\documentclass[12pt]{iopart}

\usepackage{iopams}
\usepackage{graphicx,amssymb}
\usepackage{color}

\newtheorem{proposition}{Proposition}
\newtheorem{theorem}{Theorem}

\newtheorem{definition}{Definition}

\newcommand{\be}{\begin{equation}}
\newcommand{\ee}{\end{equation}}
\def\dif{{\rm d}}
\def\ci{\mathop{\textrm{i}}\nolimits}

\def\Ric{{\rm Ric}}
\def\r{{\rm r}}

\def\ci{{\rm i}}

\begin{document}
\title[Dimension of the isometry group in type N vacuum solutions]
{Dimension of the isometry group in type N vacuum solutions: an IDEAL approach}

\author{Juan Antonio S\'aez$^1$, Salvador Mengual$^{2}$\footnote{Author to whom any correspondence should be adressed.} and Joan Josep Ferrando$^{2,3}$}

\address{$^1$\ Departament de Matem\`atiques per a l'Economia i l'Empresa,
Universitat de Val\`encia, E-46022 Val\`encia, Spain}

\address{$^2$\ Departament d'Astronomia i Astrof\'{\i}sica, Universitat
de Val\`encia, E-46100 Burjassot, Val\`encia, Spain}

\address{$^3$\ Observatori Astron\`omic, Universitat
de Val\`encia, E-46980 Paterna, Val\`encia, Spain}

\ead{juan.a.saez@uv.es; salvador.mengual@uv.es; joan.ferrando@uv.es}

\begin{abstract}
The necessary and sufficient conditions for a type N vacuum solution (with cosmological constant) to admit a group of isometries of dimension $r$ are given in terms of the invariant concomitants of the Weyl tensor. This study requires defining several invariant classes, and for each class, the conditions that determine the dimension are given. Thus, an IDEAL (Intrinsic, Deductive, Explicit and ALgorithmic) characterisation of these spacetimes follows. Some examples show that our algorithmic results can be easily implemented on the {\em xAct Mathematica} suite of packages. The relation between our classes and already known families of solutions of Einstein equations is outlined.
\end{abstract}
%

%

\today

\section{Introduction}

The study of the isometry group admitted by a spacetime is a significant issue in General Relativity for its interest in determining solutions of the Einstein equations and in understanding the geometric properties of the gravitational field. The problem of finding all the Killing vectors admitted by a spacetime or getting a set of invariants characterising the number of such symmetries is an old well-established problem in an $n$-dimensional Riemannian space (see for example the text by Eisenhart \cite{eisenhart-33}), and many efforts have been devoted to analyse this topic in Relativity (see for example \cite{kramer}).

The invariant method developed by Cartan \cite{cartan} to identify a Riemannian geometry was adapted to a four-dimensional spacetime by Brans \cite{brans}. In the eighties, Karlhede \cite{karlhede} and other authors (see also \cite{karlhede-maccallum}) spread this approach and applied it to the metric equivalence problem. A lot of work has been done on this topic since then (see \cite{MacCallum-2015} and references therein). 

Based on the Eisenhart \cite{eisenhart-33} and Defrise \cite{Defrise} results, the Cartan-Karlhede method can also be used to obtain the dimension of the orbits and the number of  Killing fields admitted by a given spacetime \cite{kramer, karlhede-maccallum}. Although this method is theoretically clearly established, the number of scalars that have to be computed is formidable (see \cite{tomoda} and references therein for a recent review on this subject), and general algorithms implementing this approach are difficult to build. 

In \cite{SMF-ST} we have performed an alternative approach that can be applied when the spacetime admits a {\em Riemann frame} ($R$-frame), that is, a frame that can be built from the Riemann tensor and its covariant derivatives. Our method enables us to get the dimension of the group of isometries in an algorithmic manner by taking the successive derivatives up to the fourth order of the so called
{\it connection tensor}. An analogous scheme has been used in obtaining the dimension of the isometry group in three-dimensional Riemannian spaces \cite{FS-K3}, and in studying the homogeneous three-dimensional  spaces \cite{FS-G3, FS-L3}.

Spacetimes of Petrov-Bel types I, II and III always admit a Riemann frame, which can be explicitly obtained from the Weyl tensor \cite{FMS-Weyl}. Consequently, the problem of finding the dimension of the isometry group for a solution of any of those types is fully solved in \cite{SMF-ST}. The algorithm obtained in that study is independent of the Ricci tensor. 

For solutions of Petrov-Bel types D, N or O the existence of a $R$-frame depends on the algebraic properties of the Ricci tensor or on the successive covariant derivatives of both the Ricci and Weyl tensors. Consequently, the determination of the number of isometries for these types requires a wider analysis. 

For example, we have recently characterised the spatially-homogeneous cosmologies in \cite{SMF-SHC}. In this case, the Ricci tensor defines an invariant time-like direction, and then the type N solutions admit a $R$-frame determined algebraically from both the Weyl and the Ricci tensors \cite{SMF-ST}. The study of the type D solutions demands considering many cases, some of which admit a Riemann frame defined by the Weyl and Ricci covariant derivatives \cite{SMF-SHC}. 

Therefore, the full study of the isometry group dimension in spacetimes of types D, N or O is still an open problem. In this paper, we present a first step in solving this question: the study of this problem for the type N vacuum solutions with cosmological constant $\Lambda$. Note that, in this case, the Ricci tensor does not define any scalar or tensorial invariant, and consequently, all the characterisation conditions depend, necessarily, on the Weyl tensor and its covariant derivatives. 

Our study requires considering an invariant classification of the type N vacuum solutions. The {\em regular classes} are defined by a condition that ensures the existence of a $R$-frame. For them, we can apply the results in \cite{SMF-ST}. For the {\em singular classes} (no $R$-frame exists), we need to know the scalars, directions and 2-forms that can be obtained from the Weyl tensor, and we make use of the Eisenhart \cite{eisenhart-33} and Defrise \cite{Defrise} results to determine the dimension of the isometry group and of its orbits. All this analysis offers an IDEAL (Intrinsic, Deductive, Explicit and ALgorithmic) labelling of each of the considered classes, which allows us to build algorithms that can easily be implemented on the current tensor calculus packages.

The specetimes of Petrov-Bel type N, and in particular the vacuum solutions, have been widely considered in the literature, and significant families of solution are known \cite{kramer, Griffiths}. In most cases, the null fundamental direction $\ell$ of the Weyl tensor defines a shear-free geodesic congruence. In the family of the Robinson-Trautman metrics \cite{Robinson-Trautman}, $\ell$ is non-twisting and diverging, while the non-twisting and non-diverging conditions characterise the Kundt family of solutions \cite{Kundt, Ehlers-Kundt}. A first analysis and classification of the vacuum Kundt solutions with cosmological constant was presented in \cite{Garcia-Plebanski}. These results were completed and extended to a radiation source in \cite{Oz-Ro-Ro}, and subsequently many papers have been devoted to go deeper into this subject (see, for example \cite{Bikac-Podolsky-1999, Ortaggio}). Some significant subfamilies of the vacuum Kundt solutions have been analysed using the Cartan-Karlhede or the scalar differential invariants approaches (see \cite{Coley-2013, Kruglikov} and references therein).

Despite this broad bibliography on the type N metrics, the study of the isometry group admitted by a given type N solution is still an open problem which we try to partially solve here. However, we must mention that the isometries of the vacuum pp-waves (Kundt metrics admitting a covariantly constant null vector) were studied years ago by Ehlers and Kundt \cite{Ehlers-Kundt}, and Sippel and Goenner  expanded this study to the non-vacuum case in a subsequent paper \cite{Sippel-Goenner-1986}. The symmetries of some families of non-twisting vacuum solutions (with $\Lambda$) were addressed by Salazar {\it et al.} \cite{Salazar-Garcia-Plebanski}, and McIntosh \cite{McIntosh} summarises results about symmetries in both twisting and non-twisting vacuum metrics. 
The symmetries of the Siklos solutions (a subfamily of the Kundt metrics with negative cosmological constant) have also been analysed \cite{Siklos}.

It is worth remarking on the peculiar characteristics of our IDEAL approach with respect to the usual studies of the isometry group. In the above quoted papers, the authors start from a precise canonical form of the metric tensor and study the constraints that the existence of a specific isometry group imposes on the metric functions. But these analyses do not solve the following question: given a metric tensor in an {\em arbitrary} coordinate system, $g_{\alpha \beta}(x^{\gamma})$, when is it a type N vacuum solution (with $\Lambda$) admitting a r-dimensional group of isometries G$_r$ which acts on q-dimensional orbits O$_q$? Here, we give a complete answer to this question.   

In section \ref{sec-Riemann-frame} we briefly summarise the results in \cite{SMF-ST} that we need to develop our study. We define the connection tensor $H$ and its differential concomitants $C^{[p]}$, which enable us to state the characterisation theorem on the dimension of the isometry group when a $R$-frame exists.

Section \ref{typeN-notation} is devoted to introduce basic notation and to analyse the geometric elements defined by a type N Weyl tensor. We also explain how to build a $R$-frame when complementary invariant geometric elements depending on the Ricci tensor and on derivatives of the Weyl tensor are known.   

In section \ref{sec_classes_regulars} we analyse seven regular classes defined without any restraint on the Ricci tensor. We give their invariant definition in terms of the Weyl tensor and present how to build the $R$-frame associated with each of the classes. We also study how our classification constrains the optical scalars defined by the null fundamental direction that a type N Weyl tensor admits. Finally, the limitations on the Ricci tensor imposed by the classes are analysed. It is worth remarking that these seven studied classes include the twisting metrics, the Robinson-Trautman solutions and a wide family of the Kundt metrics, including the pp-waves with trivial isotropy group.  

The analysis of the previous section does not consider a family of type N metrics defined by significant constraints on the first and second-order Weyl derivatives. Section \ref{sec-N-lambda} is devoted to study the symmetries of this family for the vacuum solutions with cosmological constant $\Lambda$. When $\Lambda=0$ we obtain the plane waves, and if $\Lambda \not=0$ the analysis requires defining five regular classes, for which we determine a $R$-frame, and four singular classes, for which we obtain the orbits and group dimensions. A subfamily of the Siklos solutions belongs to this family.

Finally, in section \ref{sec-discussion} we address three different aspects. Firstly, we discuss about the nature of our results and the relation with other previous studies. Secondly, we consider several examples that illustrate how to implement our results on {\it xAct}, a suite of packages for {\it Mathematica}. Thirdly, we also comment on further work and open problems.


\section{Isometries in spacetimes with a Riemann frame} 
\label{sec-Riemann-frame}

Let us consider a spacetime with metric $g$ of signature $\{-+++\}$ and metric volume element $\eta$. We shall denote with the same symbol a tensor and the metric  equivalent tensors that follow by raising and lowering indexes with $g$. A summary of the notation used in this paper can be found in \ref{apendix-A}.


Let $\{ e_a \}$ be an oriented orthonormal frame, $\eta = e_0 \wedge e_1 \wedge e_2 \wedge e_3$, and $\{ \theta^{a} \}$ its algebraic dual basis. The connection coefficients $\gamma_{ab}^c $ are defined as usual, $\nabla e_{a} = \gamma_{ab}^{c} \, \theta^{b} \otimes e_{c}$. In what follows, we will use Latin indexes to count the vectors of the frame, and to indicate the components of a tensor in this non-holonomic frame, while we will use Greek indexes to indicate components in a coordinate frame.

We can collect all the connection coefficients in the connection tensor $H$ \cite{SMF-ST}:
\begin{equation} \label{defH}
H= \gamma^{c}_{a b} \, \theta^{b} \otimes \theta^a \otimes e_{c} = - \frac12 \tilde{\eta}^{a b}\nabla e_{a} \bar{\wedge} e_b  \, ,
\end{equation}
where $\tilde{\eta}^{ab}$ is the signature symbol, $\tilde{\eta}^{ab}= diag(-1,1,1,1)$, and where, for a vector $v$ and a 2-tensor $A$, $(A \bar{\wedge} v)_{\alpha \beta \gamma} = A_{\alpha \beta} v_\gamma \!- \! A_{\alpha \gamma} v_\beta$ and $(v \bar{\wedge} A)_{\beta \gamma \alpha} =  v_\beta A_{\gamma \alpha} \!- \!  v_\gamma A_{\beta \alpha}$. 

The first directional derivatives of the connection coefficients along the frame, $e_{a_1}(\gamma^a_{bc})$, and the successive 
%
%
covariant $q$-derivatives of the connection coefficients, $e_{a_q} \cdots e_{a_1} (\gamma^{c}_{a b})$, can be collected in some 
%
%
%
differential concomitants (of order $q$) of $H$, $C^{[q]} \equiv C^{[q]}(H)$ (see their explicit definition in  \cite{SMF-ST}). 
%
%



A Riemann frame ($R$-frame) is a frame that can be built from the Riemann tensor and its covariant derivatives. If a $R$-frame exists, we can always orthonormalise it and then work with such oriented orthonormal frame $\{ e_a \}$. Then, its associated connection tensor $H$ and all the tensors $C^{[q]}(H)$ are differential concomitants of the Riemann tensor.

This fact enables us to carry out a flowchart (see figure 1 in \cite{SMF-ST}) that performs an algorithm providing
the dimension of the isometry group of a spacetime admitting a $R$-frame. This flow diagram uses as initial input data the connection tensor $H$ and its differential concomitants $C^{[q]}$, $q=1,2,3,4$. 

When the connection tensor $H$ defined by a $R$-frame is obtained as an explicit concomitant of the Riemann tensor, this algorithm becomes an IDEAL (Intrinsic, Deductive, Explicit and ALgorithmic) characterisation of the geometries admitting a G$_r$ group of isometries. 

The existence of such a frame strongly depends on the invariant geometric elements associated with the curvature tensor. 
In spacetimes of Petrov-Bel types I, II or III, the Weyl tensor defines a principal frame. In \cite{FMS-Weyl} we have presented an algorithmic way to determine this Weyl-frame, and in \cite{SMF-SHC} we have shown that its associated connection tensor can also be determined without explicitly obtaining the frame.

In types O, N and D the algebraic structure of the Weyl tensor does not determine a Weyl-frame to compute $H$. Then, we need to obtain new invariant directions from the Ricci tensor or from the derivatives of the Weyl tensor. Elsewhere, we have considered several of these circumstances in studying an invariant approach to the spatially-homogeneous cosmologies \cite{SMF-SHC}.


\section{Notation and invariants in type N spacetimes}
\label{typeN-notation}

Here we use the notation adopted in \cite{FMS-Weyl}. On the six-dimensional vectorial space of the 2-forms, two different
metrics can be considered: the usual one, $G=\frac{1}{2}g \wedge g$, ($g \wedge g)_{\alpha \beta \mu \nu} = 2(g_{\alpha \mu }g_{\beta  \nu}-g_{\alpha \nu}g_{\beta \mu})$, induced by the spacetime metric, and that defined by the metric volume
element $\eta$, i.e.:
$$G(F,L) \equiv (F,L) =\frac{1}{4} G_{\alpha \beta \mu \nu}
   F^{\alpha \beta} L^{\mu \nu}\ ;\ \ \ \
 \eta (F,L) \equiv (*F,L)= \frac{1}{4}
\eta_{\alpha \beta \mu \nu} F^{\alpha \beta} L^{\mu \nu} .$$
The scalar invariants of a 2-form $F$ are defined by its squares
calculated with these two metrics: $(F,F)$ and $(*F,F)$.
A 2-form is  named null when  $(F,F) = (*F,F)=0$. Simple 2-forms are those that satisfy $(*F,F) = 0$.

A null 2-form $L$ and its dual $*L$ admit a unique common (null)
eigendirection $\ell$ with zero associated eigenvalue: $L(\ell)=0$, $*L(\ell)=0$. There exists only one parametrisation of $\ell$ such that it is future-pointing and $L =\ell \wedge e_2$, where $e_2$ is a space-like unitary vector orthogonal to $\ell$, and fixed up to the change $e_{2}\hookrightarrow{e_{2}+\zeta {\ell}}$. With this canonical parametrisation we name $\ell$ the {\it fundamental vector} of $L$. 

We will denote $U$ a timelike unitary simple 2-form, ($U,*U)=0$, $(U,U)=-1$, which for short we name {\em t-volume} since it is the volume element of a time-like 2-plane.

A self-dual 2-form is a complex 2-form ${\cal F}$ such that
$*{\cal F}= \textrm{i}{\cal F}$. There exists a one-to-one correspondence between every real 2-form $F$ and the self-dual 2-form ${\cal F}=\frac{1}{\sqrt{2}}(F-\textrm{i}*F)$. We refer here to a self-dual 2-form as a {\it bivector}.

For every bivector ${\cal F}$, we have $({\cal F},{\cal F})=(F,F)-\textrm{i}(*F,F)$. Thus, a  bivector ${\cal L}=\frac{1}{\sqrt{2}}(L-\textrm{i}*L)$ is a null vector for $G$ when $L$ is a null 2-form, and a $G$-unitary bivector ${\cal U}=\frac{1}{\sqrt{2}}(U-\textrm{i}*U)$ corresponds to every t-volume $U$. 

If ${\cal F}$ is a non-null bivector, then $F$ admits an invariant time-like plane. The corresponding t-volume can be obtained as:
\be \label{t-volume}
U = \frac{1}{\sqrt{2}}({\cal U} + \tilde{{\cal U}}) , \qquad     {\cal U} = \frac{{\cal F}}{\sqrt{-({\cal F} ,{\cal F})}}  .
\ee
If $F$ is a simple and time-like 2-form ($(F, *F)=0$, $(F, F) <0$), the associated t-volume is given by:
\be \label{t-volume_2}
U = \frac{F}{\sqrt{-(F ,F)}}  .
\ee
%


\subsection{Invariant algebraic elements associated with a type N Weyl tensor}
\label{geometry_typeN}

If ${\cal W}$ is the self-dual Weyl tensor, ${\cal W}= \frac{1}{2} (W - \ci *W)$, we denote ${\cal W}^2$ the double bivector $({\cal W}^2)_{\alpha \beta \mu \nu} = \frac{1}{2} {\cal W}_{\alpha \beta \lambda \rho} {{\cal W}^{\lambda \rho}}_{\mu \nu}$. A type N Weyl tensor is characterised by the condition ${\cal W}^2 = 0$. Then, a null bivector ${\cal L}$ exists such that 
\be \label{Weyl-N}
{\cal W}={\cal L} \otimes {\cal L} , \qquad W = L \otimes L - *L \otimes *L .
\ee
Let $\ell$ be the fundamental vector of the bivector ${\cal L}$. Then, ${\cal L}$ and $\ell$ are the {\em fundamental bivector}  and the {\em fundamental vector} of a type {\em N} Weyl tensor. We have \cite{FMS-Weyl}:
\begin{proposition} \label{prop-L_l}
The fundamental bivector ${\cal L}$ and the fundamental vector $\ell$ of a type {\em N} Weyl tensor (${\cal W}={\cal L} \otimes {\cal L}$, $\ell \wedge {\cal L} =0$) can be obtained as:
\begin{equation} \label{L-ell}
{\cal L} \equiv \frac{{\cal W}({\cal X})}{\sqrt{{\cal W}({\cal X},{\cal X})}} \ ,\quad  \quad \ell \equiv \frac{L^2(x)}{\sqrt{-L^2(x,x)}}\, , \quad L \equiv \frac{1}{\sqrt{2}}({\cal L} + \tilde{{\cal L}}) \, ,
\end{equation}
where ${\cal X}$ is an arbitrary bivector such that ${\cal W}({\cal X}) \not= 0$ and $x$ an arbitrary time-like vector.
\end{proposition}
Note that both ${\cal L}$ and $\ell$ are determined up to sign, but (\ref{L-ell}) gives a future-pointing fundamental vector $\ell$ if $x$ is also future-pointing.

It is worth remarking that the algebraic invariant elements $\ell$ and ${\cal L}$ defined by a type N Weyl tensor do not determine a $R$-frame. Nevertheless, it is useful to consider frames partially determined by these invariant elements for our analysis. A null {\em adapted frame} of a type N Weyl tensor $\{\ell, k, e_2, e_3\}$ is defined by the following conditions: $\ell$ is the fundamental vector, $L = \ell \wedge e_2$, $*L = -\ell \wedge e_3$, $L$ being the fundamental 2-form, and $(\ell, k)=-1$. An adapted frame is fixed up to the changes $e_{2}\hookrightarrow{e_{2}+\zeta {\ell}}$, $e_{3}\hookrightarrow{e_{3}+\zeta^* {\ell}}$, $k \hookrightarrow k +\zeta e_{2}+\zeta^* e_{3}$, where $\zeta$ and $\zeta^*$ are arbitrary scalars. In what follows, ${\cal U}$ denotes the unitary bivector associated with the t-volume $U = \ell \wedge k$. Note that $({\cal L}, {\cal U})=0$.


\subsection{Riemann frames in type N spacetimes}
\label{R-frames_N}

From now on, by {\em invariant} tensor (scalar, vector,...) we refer to a tensor that can be obtained from the Riemann tensor and its covariant derivatives. 

\begin{proposition} \label{prop-U_base}
Let ${\cal L}$ and $\ell$ be the fundamental bivector and the fundamental vector of a type {\em N} Weyl tensor. If $U$ is an invariant t-volume which is orthogonal to ${\cal L}$ (${\cal L}, U)=0$ ($\leftrightarrow \ell \wedge U(\ell) =0$), then a null $R$-frame $\{\ell, k, e_2, e_3\}$ exists, which is given by :
\begin{equation}
\hspace{-22mm} k = -\frac{(U^2\!+\!\epsilon U)(x)}{2 (x, \ell)} , \ \ \epsilon \! =\!\frac{U(x,\ell)}{(x, \ell)}, \quad e_2\! =\! L(k) , \ \ e_3 \!=\! - *\!L(k)  , \ \ L\! =\! \frac{1}{\sqrt{2}}({\cal L}\! +\! \tilde{{\cal L}}) \, . 
\end{equation}
where $x$ is an arbitrary time-like vector.
\end{proposition}
\begin{proposition} \label{prop-{b,c}_base}
Let  $\ell$ be the fundamental vector of a type {\em N} Weyl tensor. 
If $\{b,c\}$ are two invariant vectors such that $(\ell, b)= (\ell, c)= 0$ and $\ell \wedge b \wedge c \not =0$, then a null $R$-frame $\{\ell, k, e_2, e_3\}$ exists, which is given by :
\begin{eqnarray}
k = -\frac{(U^2+ \epsilon U)(x)}{2 (x, \ell)} , \qquad \epsilon =\frac{U(x,\ell)}{(x, \ell)}, \qquad U = *(e_2 \wedge e_3) ,\\  e_2 = \frac{b}{\sqrt{b^2}} , \qquad e_3= \frac{\hat{c}}{\sqrt{\hat{c}^2}}, \qquad \hat{c} = c- \frac{(c,b)}{b^2}b ,  
\end{eqnarray}
where $x$ an arbitrary time-like vector.
\end{proposition}
The null frame $\{\ell, k, e_2, e_3\}$ obtained in proposition \ref{prop-U_base} is oriented, and the one obtained in proposition \ref{prop-{b,c}_base} becomes oriented with the change $e_3 \hookrightarrow \epsilon e_3$. From these null frames we can determine an orthonormal $R$-frame $\{e_0, e_1, e_2, e_3\}$, $\sqrt{2}\, e_0 = \ell + k , \ \sqrt{2}\, e_1 = \ell - k$, and obtain the associated connection tensor using (\ref{defH}). Alternatively, we can use the following expression that we easily obtain from (\ref{defH}):  
\begin{equation} \label{defH_null}
2 H= \nabla \ell\, \bar{\wedge}\, k + \nabla k \, \bar{\wedge} \, \ell -   \nabla e_{2} \bar{\wedge}\, e_2 -  \nabla e_{3} \bar{\wedge}\, e_3 .
\end{equation}
%


\section{Some regular classes of type N spacetimes}
\label{sec_classes_regulars}

Here we introduce several differential concomitants of the Weyl tensor that enable us to define some regular classes of type N spacetimes. For each of these classes we determine the corresponding $R$-frame. To better understand our reasoning, we offer the expression (placed in brackets) of these concomitants in an adapted frame using the structure equations given in \ref{apendix-A}.   

Let us consider the first-order differential concomitant of the Weyl tensor given by: 
\be \label{F_1}
{\cal F}_1 \equiv \nabla_{\ell}{\cal L} \quad \big[= (\ell, w- \ci v) \ {\cal L} + (\ell, a- \ci a^*) \ {\cal U} \big],
\ee
which is a bivector orthogonal to ${\cal L}$.
If ${\cal F}_1$ is a non-null bivector, (${\cal F}_1,{\cal F}_1) \not=0$, then we can determine a t-volume $U_1$ as in (\ref{t-volume}) with ${\cal F} = {\cal F}_1$. This t-volume fulfils the conditions of proposition \ref{prop-U_base} and a $R$-frame can be obtained.

Let us define the invariant vectors:
\be \label{a_a*}
a \equiv -\frac{1}{(\ell,x)}(\nabla L \cdot \ell)(x) , \qquad a^* \equiv \frac{1}{(\ell,x)}(\nabla \! *\! L \cdot \ell)(x) ,
\ee
where $x$ is an arbitrary time-like vector. When ${\cal F}_1$ is a null bivector, $({\cal F}_1,{\cal F}_1)=0$, these vectors are orthogonal to $\ell$, $(a, \ell)= (a^*, \ell)=0$ (see (\ref{F_1})). Then, we can define the first-order differential concomitants of the Weyl tensor:
\be \label{F_2}
\begin{array}{l}
{\cal F}_2 \equiv \nabla_{a}{\cal L} \ \quad \big[= (a, w- \ci v) \ {\cal L} + [a^2 + \ci (a, a^*)] \ {\cal U} \big], \\[1mm]  {\cal F}_2^* \equiv \nabla_{a^*}{\cal L} \quad   \big[= (a^*, w- \ci v) \ {\cal L} + [(a, a^*) + \ci a^{*2}] \ {\cal U} \big], 
\end{array}
\ee
which are bivectors orthogonal to ${\cal L}$.
If ${\cal F}_2$ (respectively, ${\cal F}_2^*$) is a non-null bivector, $({\cal F}_2,{\cal F}_2) \not=0$ (respectively $({\cal F}_2^*,{\cal F}_2^*) \not=0$), then we can determine a t-volume $U_2$ (respectively, $U_2^*$) as in (\ref{t-volume}) with ${\cal F} = {\cal F}_2$ (respectively, ${\cal F} = {\cal F}_2^*$). This t-volume fulfils the conditions of proposition \ref{prop-U_base} and a $R$-frame can be obtained.

When $({\cal F}_1,{\cal F}_1)=({\cal F}_2,{\cal F}_2) = ({\cal F}_2^*,{\cal F}_2^*) =0$, then $a = \alpha \ell$, $a^* = \alpha^* \ell$ (see (\ref{F_2})), and the invariant scalars $\alpha$ and $\alpha^*$ can be obtained as:
\be \label{alpha}
\alpha \equiv \frac{(a,x)}{(\ell,x)} , \qquad  \alpha^* \equiv \frac{(a^*,x)}{(\ell,x)} ,
\ee
where $x$ is an arbitrary time-like vector. If $^t\!A$ denotes the transposed of a 2-tensor, $(^t\!A)_{\alpha \beta} = A_{\beta \alpha}$, we have now: 
\be \label{B}
\nabla \ell + ^t \! \nabla \ell \equiv  B = \ell \, \tilde{\otimes}\, b ,
\ee
where $b$ is an invariant vector that can be obtained as:
\begin{equation} \label{b-w}
b \equiv \frac{1}{(\ell,x)} \left[B(x) - \frac{B(x,x)}{2 (\ell, x)} \ell \right]   \ \quad \big[= w + \alpha e_2 + \alpha^* e_3 \big]   , 
\end{equation}
where $x$ is an arbitrary time-like vector. Let us define the first-order differential concomitant of the Weyl tensor:
\be \label{F_3}
F_3 \equiv \dif \ell  \quad \big[=  w \wedge \ell + \ell \wedge (\alpha e_2 + \alpha^* e_3) \big] ,
\ee
which is a simple 2-form ($(F_3, *F_3)=0$) orthogonal to ${\cal L}$. 
If $F_3$ is a non-null 2-form, then it is time-like, $(F_3, F_3) <0$, and we can determine a t-volume $U_3$ as in (\ref{t-volume_2}) with $F=F_3$. This t-volume fulfils the conditions of proposition \ref{prop-U_base} and a $R$-frame can be obtained.

When $({\cal F}_1,{\cal F}_1)=({\cal F}_2,{\cal F}_2) = ({\cal F}_2^*,{\cal F}_2^*) =(F_3, F_3)=0$, we have $(\ell, b)= 0$ as a consequence of (\ref{b-w}) and (\ref{F_3}). Then, we can define the invariant 3-tensor:
\be \label{C} 
C \equiv - Sim(\nabla\!  *\!L \cdot L) = [\nabla_{(\alpha}\! *\!L_{\beta}^{\ \mu} L_{\gamma) \mu}] .
\ee
A straightforward calculation leads to 
$3 C = \ell \otimes \ell \otimes c + \ell \otimes c \otimes \ell + c \otimes \ell \otimes \ell$, $c$ being an invariant vector that can be obtained as: 
\be \label{c-v}
c \equiv \frac{1}{(\ell,\!x)^3}[3(\ell,\! x) C(x,\!x)\! -\! 2 C(x,\!x,\!x) \ell]   \ \quad \big[= v + \alpha e_3 - \alpha^* e_2 \big]   ,
\ee
where $x$ is an arbitrary time-like vector. 
Let us define the first-order differential concomitant of the Weyl tensor:
\be \label{F_4}
F_4 \equiv \ell \wedge c,
\ee
which is a simple 2-form ($(F_4, *F_4)=0$) orthogonal to ${\cal L}$. If $F_4$ is a non-null 2-form, then it is time-like, $(F_4, F_4) <0$, and we can determine a t-volume $U_4$ as in (\ref{t-volume_2}) with $F=F_4$. This t-volume fulfils the conditions of proposition \ref{prop-U_base} and a $R$-frame can be obtained.

When $({\cal F}_1,{\cal F}_1)=({\cal F}_2,{\cal F}_2) = ({\cal F}_2^*,{\cal F}_2^*) =(F_3, F_3)=(F_4, F_4)=0$, we have $(\ell, c)=0$. Then, we can define the second-order differential concomitant of the Weyl tensor:
\be \label{F_6}
F_5 \equiv \dif b  ,
\ee
which, as we will see in (\ref{db-dc}), is a simple 2-form ($(F_5, *F_5)=0$) orthogonal to ${\cal L}$. If $F_5$ is a non-null 2-form, then it is time-like, $(F_5, F_5) <0$, and we can determine a t-volume $U_5$ as in (\ref{t-volume_2}) with $F=F_5$. This t-volume fulfils the conditions of proposition \ref{prop-U_base} and a $R$-frame can be obtained.

When $({\cal F}_1,{\cal F}_1)=({\cal F}_2,{\cal F}_2) = ({\cal F}_2^*,{\cal F}_2^*) =(F_3, F_3)=(F_4, F_4)= (F_5, F_5)= 0$, we can define the second-order differential concomitants of the Weyl tensor:
\be \label{f_f*}
f \equiv \dif \alpha -\alpha^* c, \qquad f^* \equiv \dif \alpha^* +\alpha c,
\ee
which are two vectors orthogonal to $\ell$, $(\ell, f) = (\ell, f^*) =0$ as we will see in (\ref{d-alpha}). If $\varphi_6 \equiv *(\ell \wedge f \wedge f^*) \not=0$, then the vectors $\{f,f^*\}$ fulfil the conditions of proposition \ref{prop-{b,c}_base} and a $R$-frame can be obtained.

When $({\cal F}_1,{\cal F}_1)=({\cal F}_2,{\cal F}_2) = ({\cal F}_2^*,{\cal F}_2^*) =(F_3, F_3)=(F_4, F_4)= (F_5, F_5)= 0$ and $\varphi_6=0$, we can define $\varphi_7 \equiv *(\ell \wedge b \wedge c)$. If $\varphi_7 \not=0$, then $\{b,c\}$ fulfil the conditions of proposition \ref{prop-{b,c}_base} and a $R$-frame can be obtained.


\subsection{Summary: classification and characterisation theorem}
\label{Summary}

The analysis presented above induces a classification of the studied type N spacetimes. Now we introduce the definition and notation of each class, which is useful to summarise the results and to make the presentation of the rest of the paper clearer.
\begin{definition} \label{def-Cn} 
Let ${\cal L}$ and $\ell$ be the fundamental bivector and the fundamental vector of a type {\em N} Weyl tensor. Let us define the Weyl concomitants:
\be
\begin{array}{l}
\hspace{-15mm} \varphi_1 \equiv ({\cal F}_1,{\cal F}_1), \quad \varphi_2 \equiv |({\cal F}_2,{\cal F}_2)| + |({\cal F}_2^*,{\cal F}_2^*)|,  \quad \varphi_3 \equiv (F_3, F_3), \\    \hspace{-15mm} \varphi_4 \equiv (F_4, F_4,),   \quad \varphi_5 \equiv (F_5, F_5), \quad \varphi_6 \equiv *(\ell \wedge f \wedge f^*) , \quad \varphi_7 \equiv *(\ell \wedge b \wedge c) ,
\end{array}
\ee
where ${\cal F}_1$ is given in {\em (\ref{F_1})},  ${\cal F}_2$ and ${\cal F}_2^*$ are given in {\em (\ref{a_a*})} and {\em (\ref{F_2})}, $F_3$ is given in {\em (\ref{F_3})}, $F_4$ is given in {\em (\ref{F_4})}, $b$ is given in {\em (\ref{B})} and {\em (\ref{b-w})}, $c$ is given in {\em (\ref{C})} and {\em (\ref{c-v})}, $F_6$ is given in {\em (\ref{F_6})} and $f$ and $f^*$ are given in {\em (\ref{alpha})} and {\em (\ref{f_f*})}.
\begin{itemize}
\item[$\circ$]
We say that a type N spacetime is of class {\rm C}n if $\varphi_i =0$ $\forall i < n$ and $\varphi_n \not= 0$. 
\item[$\circ$]
We denote $\widehat{{\rm C}n}$ the family of type N spacetimes such that $\varphi_i =0$ $\forall i \leqslant  n$.
\end{itemize}
\end{definition}
All the Weyl invariants $\varphi_n$ defining the classes {\rm C}n are scalars except $\varphi_6$ and $\varphi_7$ which are vectors. The invariants $\varphi_n$, $n \leqslant 4$, and $\varphi_7$ are of first-order in the Weyl tensor, while $\varphi_5$ and $\varphi_6$ are of second-order. Note that $\widehat{{\rm C}n}$ is the complementary of $\bigcup_{p=1}^{n} {\rm C}p$. 
\begin{theorem} \label{theorem-Cn}
All the type N spacetimes of class {\rm C}n, $n= 1,...,7$, are regular, that is, they admit a $R$-frame ${\cal B}_n =  \{\ell, k, e_2, e_3\}$, which can be obtained as follows:
\begin{itemize}
\item[{\rm C1}]
The $R$-frame is obtained as stated in proposition {\em \ref{prop-U_base}}, with the t-volume $U$ given by {\em (\ref{t-volume})}, with ${\cal F} = {\cal F}_1$. 
\item[{\rm C2}]
The $R$-frame is obtained as stated in proposition 
{\em \ref{prop-U_base}}, with the t-volume $U$ given by {\em (\ref{t-volume})}, with ${\cal F} = {\cal F}_2$ if $({\cal F}_2, {\cal F}_2) \not=0$ or ${\cal F} = {\cal F}_2^*$ if $({\cal F}_2^*, {\cal F}_2^*) \not=0$.
\item[{\rm C3}]
The $R$-frame is obtained as stated in proposition {\em \ref{prop-U_base}}, with the t-volume $U$ given by {\em(\ref{t-volume_2})}, with $F = F_3$. 
\item[{\rm C4}]
The $R$-frame is obtained as stated in proposition {\em \ref{prop-U_base}}, with the t-volume $U$ given by {\em(\ref{t-volume_2})}, with $F = F_4$. 
\item[{\rm C5}]
The $R$-frame is obtained as stated in proposition {\em \ref{prop-U_base}}, with the t-volume $U$ given by {\em(\ref{t-volume_2})}, with $F = F_5$. 
\item[{\rm C6}]
The $R$-frame is obtained as stated in proposition {\em \ref{prop-{b,c}_base}}, with $\{b, c\} = \{f, f^*\}$. 
\item[{\rm C7}]
The $R$-frame is obtained as stated in proposition {\em \ref{prop-{b,c}_base}}. 
\end{itemize}
If $H_n$ is the connection tensor {\em (\ref{defH_null})} associated  with ${\cal B}_n$, the dimension of the isometry group is obtained as stated in the algorithm given in figure {\em 1} of {\em \cite{SMF-ST}}.
\end{theorem}

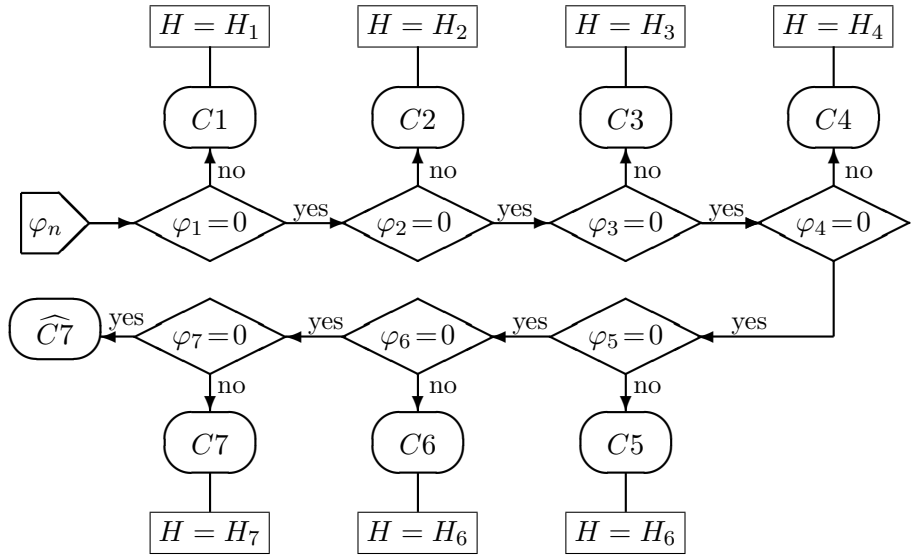
\begin{figure}[h]
\vspace*{-3.2cm} 

 \setlength{\unitlength}{1cm} {\small \noindent
 \hspace*{10mm}
\begin{picture}(1,13)


\thicklines


 \put(1.4,6.9){\vector(1,0){0.6}}
\put(4,6.9){\vector(1,0){0.8}} \put(6.7,6.9){\vector(1,0){0.8}}
\put(9.5,6.9){\vector(1,0){0.8}}

 \put(2,5.4){\vector(-1,0){0.46}}
\put(4.8,5.4){\vector(-1,0){0.8}} \put(7.5,5.4){\vector(-1,0){0.8}}
\put(9.5,6.9){\vector(1,0){0.8}}



\put(3,4.9){\vector(0,-1){0.5}} \put(3,3.6){\line(0,-1){0.52}}

\put(5.75,4.9){\vector(0,-1){0.5}} \put(5.75,3.6){\line(0,-1){0.52}}

\put(8.5,4.9){\vector(0,-1){0.5}} \put(8.5,3.6){\line(0,-1){0.52}}


\put(3,7.4){\vector(0,1){0.5}} \put(3,8.7){\line(0,1){0.54}}

\put(5.75,7.4){\vector(0,1){0.5}} \put(5.75,8.7){\line(0,1){0.54}}

\put(8.5,7.4){\vector(0,1){0.5}} \put(8.5,8.7){\line(0,1){0.54}}

\put(11.25,7.4){\vector(0,1){0.5}} \put(11.25,8.7){\line(0,1){0.54}}

\put(11.25,6.4){\line(0,-1){1}} \put(11.25,5.4){\vector(-1,0){1.8}}


\put(0.5,6.5){\line(0,1){0.8}} \put(0.5,7.3){\line(1,0){0.5}}
\put(0.5,6.5){\line(1,0){0.5}} \put(1,7.3){\line(1,-1){0.4}}
 \put(1,6.5){\line(1,1){0.4}}
 \put(0.6,6.8){$\varphi_n$}
 \put(1.4,6.9){\line(1,0){0.6}}

\put(2.5,6.8){$\varphi_1\!=\!0$}

\put(3,7.4){\line(-2,-1){1}} \put(3,7.4){\line(2,-1){1}}
\put(3,6.4){\line(2,1){1 }} \put(3,6.4 ){\line(-2,1){1 }}

\put(5.2,6.8){$\varphi_2\!=\!0$}

\put(5.75,7.4){\line(-2,-1){1}} \put(5.75,7.4){\line(2,-1){1}}
\put(5.75,6.4){\line(2,1){1 }} \put(5.75,6.4 ){\line(-2,1){1 }}

\put(8.0,6.8){$\varphi_3\!=\!0$} \put(8.5,7.4){\line(-2,-1){1}}
\put(8.5,7.4){\line(2,-1){1}} \put(8.5,6.4){\line(2,1){1 }}
\put(8.5,6.4 ){\line(-2,1){1 }}

\put(10.75,6.8){$\varphi_4\!=\!0$} \put(11.25,7.4){\line(-2,-1){1}}
\put(11.25,7.4){\line(2,-1){1}} \put(11.25,6.4){\line(2,1){1 }}
\put(11.25,6.4 ){\line(-2,1){1 }}


\put(2.5,5.3){$\varphi_7\!=\!0$}

\put(3,5.9){\line(-2,-1){1}} \put(3,5.9){\line(2,-1){1}}
\put(3,4.9){\line(2,1){1 }} \put(3,4.9 ){\line(-2,1){1 }}

\put(5.25,5.3){$\varphi_6\!=\!0$}

\put(5.75,5.9){\line(-2,-1){1}} \put(5.75,5.9){\line(2,-1){1}}
\put(5.75,4.9){\line(2,1){1 }} \put(5.75,4.9 ){\line(-2,1){1 }}

\put(8.0,5.3){$\varphi_5\!=\!0$} \put(8.5,5.9){\line(-2,-1){1}}
\put(8.5,5.9){\line(2,-1){1}} \put(8.5,4.9){\line(2,1){1 }}
\put(8.5,4.9 ){\line(-2,1){1 }}

\put(0.95,5.5){{\oval(1.2,0.8)}}

\put(0.7,5.3){$\widehat{C7}$}


\put(3,8.3){{\oval(1.2,0.8)}} \put(2.75,8.15){$C1$}
\put(2.2,9.4){\framebox{$H=H_1$}}

\put(5.75,8.3){{\oval(1.2,0.8)}} \put(5.5,8.15){$C2$}
\put(4.95,9.4){\framebox{$H=H_2$}}

\put(8.5,8.3){{\oval(1.2,0.8)}} \put(8.25,8.15){$C3$}
\put(7.7,9.4){\framebox{$H=H_3$}}

\put(11.25,8.3){{\oval(1.2,0.8)}} \put(11,8.15){$C4$}
\put(10.45,9.4){\framebox{$H=H_4$}}


\put(3,4){{\oval(1.2,0.8)}} \put(2.75,3.85){$C7$}
\put(2.2,2.7){\framebox{$H=H_7$}}

\put(5.75,4){{\oval(1.2,0.8)}} \put(5.5,3.85){$C6$}
\put(4.95,2.7){\framebox{$H=H_6$}}

\put(8.5,4){{\oval(1.2,0.8)}} \put(8.25,3.85){$C5$}
\put(7.7,2.7){\framebox{$H=H_6$}}


 \put(4.1,7){\footnotesize{yes}}

\put(6.8,7){\footnotesize{yes}}

\put(9.6,7){\footnotesize{yes}}

\put(9.9,5.5){\footnotesize{yes}}

\put(7,5.5){\footnotesize{yes}}

\put(4.3,5.5){\footnotesize{yes}}

\put(1.66,5.55){\footnotesize{yes}}

\put(3.1,7.5){\footnotesize{no}}

\put(5.9,7.5){\footnotesize{no}}

\put(8.6,7.5){\footnotesize{no}}

\put(11.4,7.5){\footnotesize{no}}

\put(8.6,4.65){\footnotesize{no}} \put(5.9,4.65){\footnotesize{no}}

\put(3.1,4.65){\footnotesize{no}}


\end{picture} 
}

\vspace*{-2.4cm} 
\caption{This flow diagram distinguishes the regular classes C$n$, $n \leqslant 7$ and the complementary family $\widehat{\rm C7}$.} \label{figure-1}
\end{figure}

These results offer an IDEAL labelling of several regular classes of type N metrics and explain how to determine a $R$-frame for each class. This enables us to build an algorithm that we present as a flow diagram in Figure \ref{figure-1}. The sole input data are the Weyl tensor and its differential concomitants $\varphi_n $ given in definition \ref{def-Cn}. The first seven end arrows lead to the seven classes Cn and to the determination of the associated connection tensor, to which we must apply the algorithm  given in \cite{SMF-ST}. The last end arrow leads to the algorithm in figure \ref{figure-2} where the family $\widehat{\rm C7}$ is considered. 


\subsection{Some suitable remarks}
\label{Remarks}

{\bf Remark 1:} Note that the invariants used in defining the classes Cn depend on the sole Weyl tensor. The same occurs for the $R$-frames, the associated connection tensors $H_n$ and, consequently, for all the invariant tensors $C^{[p]}$ involved in the characterisation theorem \ref{theorem-Cn}. Thus, all the analysis in this section is independent of the Ricci tensor. 

Nevertheless, we will see in the following subsection that some classes restrict the Ricci tensor. And conversely, solutions with some degenerate Ricci tensors cannot belong to some of the defined classes Cn. This is the case of the Einstein spaces studied in detail in this paper (the Ricci tensor is $Ric = \Lambda g$).
\ \\[2mm]
\noindent 
{\bf Remark 2:} We have defined the classes Cn in terms of Weyl concomitants that determine the $R$-frame, when it exists. Nevertheless, there are alternative characterisations in terms of the fundamental vector $\ell$ and the fundamental bivector ${\cal L}$, which can be useful to relate our classes to the usual invariant classifications of the algebraically special solutions \cite{kramer}. In the following remarks we analyse them.
\ \\[2mm]
\noindent 
{\bf Remark 3:} The most regular class C1 can also be defined by the condition that the fundamental vector $\ell$ is not geodesic ($\kappa \not=0$ in NP formalism). In fact, we have the following equivalent conditions which characterise the family $\widehat{\rm C1}$:
\begin{equation} \label{C^1}
\varphi_1 =0 \quad \leftrightarrow \quad \nabla_{\ell} \ell \wedge \ell =0 \quad \leftrightarrow \quad (\ell,a)^2+(\ell, a^*)^2 =0 .
\end{equation}
Most of the known type N solutions have a geodesic fundamental direction (see \cite{kramer}) and, consequently, belong to $\widehat{\rm C1}$.
\ \\[2mm]
\noindent 
{\bf Remark 4:} Conditions $\varphi_1 =\varphi_2 =0$ that characterise the family $\widehat{\rm C2}$ state, equivalently, that the fundamental vector $\ell$ defines a non-twisting, non-diverging and shear-free geodesic congruence ($\kappa = \sigma =\rho =0$ in NP formalism). All these properties can be stated in a sole tensorial equation for $\ell$. Then, we have the following equivalent conditions which characterise the family $\widehat{\rm C2}$:
\begin{equation} \label{C^2}
\varphi_1 = \varphi_2 =0 \quad \leftrightarrow \quad \ell \, \bar{\wedge} \, \nabla \, \ell \, \bar{\wedge} \, \ell =0 \quad \leftrightarrow \quad \ell \wedge a = \ell \wedge a^* =0 .
\end{equation}

Then, the type N twisting solutions \cite{kramer} and the non-twisting and diverging Robinson-Trautman metrics \cite{Robinson-Trautman} belong to class C2 and, consequently, all of them admit a $R$-frame and the isotropy group is trivial.
On the other hand, the non-diverging metrics by Kundt \cite{Ehlers-Kundt, kramer} fulfil (\ref{C^2}) and they belong to $\widehat{\rm C2}$. 
\ \\[2mm]
\noindent 
{\bf Remark 5:} If (\ref{C^1}) holds, the fundamental vector $\ell$ defines a geodesic congruence. But the $\ell$ given in (\ref{L-ell}) does not correspond, necessarily, to the affine parametrisation of the congruence ($\epsilon + \bar{\epsilon}=0$, in NP formalism). The family $\widehat{\rm C3}$ is characterised by this property. More precisely, we have: 
\begin{equation} \label{C^3}
\varphi_1 = \varphi_2 =0 \quad  \Rightarrow \quad  \Big[\ \varphi_3 =0 \quad  \leftrightarrow \quad \nabla_{\ell} \ell = 0 \quad \leftrightarrow \quad  (\ell, b)=0 \ \Big].
\end{equation}
\ \\[-3mm]
\noindent 
{\bf Remark 6:} In the family $\widehat{\rm C3}$ we can always find a (non-adapted) frame such that $(\nabla_{\ell} e_2, e_3) = 0$ ($\epsilon - \bar{\epsilon}=0$, in NP formalism). When such a frame is adapted we reach the family $\widehat{\rm C4}$: 
\begin{equation} \label{C^4}
\hspace{-20mm} \varphi_1 = \varphi_2 = \varphi_3=0 \quad  \Rightarrow \quad \Big[\ \varphi_4 =0 \quad  \leftrightarrow \quad (\nabla_{\ell} e_2, e_3) = 0 \quad \leftrightarrow \quad  (\ell, c)=0 \ \Big].
\end{equation}
\ \\[2mm]
\noindent 
{\bf Remark 7:} Class C5 is specified by the condition $\varphi_5 = (\dif b, \dif b) \not=0$. Similarly, a priori, we could use the 2-form $\dif c$ to discriminate a new class. Nevertheless, it does not provide new information because $* \dif c = \dif b$ in the family $\widehat{\rm C2}$ (see following subsection). 
\ \\[2mm]
\noindent 
{\bf Remark 8:} All the classes C$n$, $n \leqslant 7$ admit a $R$-frame and, consequently, the isotropy group is trivial and the dimension of the isometry group can be determined from the algorithm given in figure 1 of \cite{SMF-ST}. Then, the metrics that we need to study are those in the  family $\widehat{\rm C7}$ for the different  compatible algebraic Ricci tensors.


\subsection{Constraints on the Ricci tensor}
\label{Ricci_constraints}

Although the above results are independent of the Ricci tensor, the Ricci and Bianchi identities link the geometric properties of the Weyl and Ricci tensors, and some of the classes introduced in definition \ref{def-Cn} are only compatible with specific algebraic types of the Ricci tensor. 

The Ricci identities for the fundamental vector $\ell$ lead to the second structure equations (\ref{2ee-2}) and (\ref{2ee-3}), which in the family $\widehat{C2}$ (constraints (\ref{C^2}), $a = \alpha \ell$, $a^* = \alpha^* \ell$), take the expressions:
\begin{equation}  \label{23ee}
\begin{array}{l} 
[\dif \alpha - \alpha^* c - (\alpha^2+ \alpha^{*2}) e_2 - Q(e_2)] \wedge \ell = e_2 \wedge Q(\ell),   \\[1mm]
[\dif \alpha^* + \alpha c - (\alpha^2+ \alpha^{*2}) e_3 - Q(e_3)] \wedge \ell = e_3 \wedge Q(\ell) ,   
\end{array}
\end{equation}
with $2 Q = Ric - \frac{r}{6} g $, $Ric$ being the Ricci tensor and $r$ its trace. 
From these expressions we obtain:
\begin{proposition} \label{prop-Q-ell}
The fundamental direction $\ell$ of the type N metric in the family $\widehat{\rm C2}$ is an eigenvector of the Ricci tensor:
\be \label{Q-ell}
Q(\ell) = \lambda \ell , \qquad \lambda \equiv \frac{Q(\ell, x)}{(\ell, x)},
\ee
where $x$ is an arbitrary timelike vector.
\end{proposition}

From (\ref{Q-ell}), equations (\ref{23ee}) become:
\begin{equation}  \label{d-alpha}
\begin{array}{l}  
\dif \alpha =  (\lambda + \alpha^2+ \alpha^{*2}) e_2 + Q(e_2) + \alpha^* c + \mu \ell ,     \\[1mm]
\dif \alpha^* =  (\lambda + \alpha^2+ \alpha^{*2}) e_3 + Q(e_3) - \alpha c + \mu^* \ell .   
\end{array}
\end{equation}
Then, from the first structure equations (\ref{1ee}) and second structure equations (\ref{2ee-1}) and (\ref{2ee-4}), and taking into account (\ref{b-w}) and (\ref{d-alpha}), we obtain: 
\begin{equation}  \label{db-dc}
\begin{array}{l}  
\dif b =  (2 \lambda + \alpha^2+ \alpha^{*2}) \ell \wedge k + \mu \ell \wedge e_2 + \mu^* \ell \wedge e_3 ,     \\[1mm]
\dif c =  (2 \lambda + \alpha^2+ \alpha^{*2}) e_2 \wedge e_3 + \mu \ell \wedge e_3 - \mu^* \ell \wedge e_2 .   
\end{array}
\end{equation}

Note that $* \dif c = \dif b$, as we have commented in the Remark 7. Moreover, they are null 2-forms ($\varphi_5 =0$) if, and only if, $2 \lambda + \alpha^2+ \alpha^{*2}=0$. When this holds, (\ref{d-alpha}) imply that the vectors defined in (\ref{f_f*}) take the expressions: 
\be \label{f-f*-Q}
f = \mu \ell -\lambda e_2 + Q(e_2)  , \qquad  f^* = \mu^* \ell -\lambda e_3 + Q(e_3) ,
\ee
which are valid  for the family $\widehat{\rm C5}$. In this case, the eigenvalue is non positive, $\lambda \leqslant 0$. Moreover, it vanishes if, and only if, $\alpha = \alpha^* =0$, which implies $f= f^* =0$ as a consequence of (\ref{f_f*}). Then, (\ref{f-f*-Q}) leads to $Q(e_2) = - \mu \ell$ and $Q(e_3) = - \mu^* \ell$. From this reasoning, we obtain:
\begin{proposition} \label{lambda-zero}
The eigenvalue $\lambda$ of $Q$ associated with the fundamental direction $\ell$ of the type N metric in the family $\widehat{\rm C5}$ fulfils:
\be \label{lambda_alpha}
2 \lambda + \alpha^2+ \alpha^{*2}=0 , \qquad  \lambda \leqslant 0, 
\ee
where $\alpha, \alpha^*$ are the Weyl invariants given in {\em (\ref{alpha})}.\\[1mm]
Moreover, $\lambda=0$ if, and only if, $\alpha = \alpha^* =0$, and then $Q = \ell \tilde{\otimes}q$, with $(\ell, q)= 0$. The spacetime is: (i) Vacuum ($Q =0$)  if $q=0$, (ii) Segre type {\em [(31)]} if  $q^2=0$, (iii) Segre type {\em [(211)]} if $q^2>0$. 
\end{proposition}

On the other hand, from (\ref{f-f*-Q}), condition $\varphi_6\not=0$ states that $\{\ell, (Q- \lambda g)(e_2), (Q- \lambda g)(e_3)\}$ is a linearly independent triplet. Consequently, for the family $\widehat{\rm C6}$ three scalars $\zeta, \nu, \nu^*$ exist such that
\be 
\zeta \ell + \nu (Q- \lambda g)(e_2) + \nu^* (Q- \lambda g)(e_3) = 0 . 
\ee
Thus, $n = \nu e_2 + \nu^* e_3$ is either an eigenvector associated with the eigenvalue $\lambda$ ($\zeta=0$), or $\{\ell, n\}$ define an invariant plane ($\zeta\not=0$). Then, we have:
\begin{proposition} \label{lambda-double}
The eigenvalue $\lambda$ of $Q$ associated with the fundamental direction $\ell$ of the type N metric in the family $\widehat{\rm C6}$ is, at least, a double eigenvalue. 
\end{proposition}
As a consequence of this result, we have several compatible Segre types of the Ricci tensor that can be distributed in two sets:
\begin{itemize}
\item[(i)]
Types [(1,1)11], [(1,1)(11)], [211], [(21)1] and  [2(11)]: the Weyl and Ricci tensors algebraically determine a $R$-frame, and the algorithm given in figure 1 of \cite{SMF-ST} can be applied.
\item[(ii)]
Types [(1,11)1], [(1,111)], [(211)], [31] and [(31)]: we need Weyl or/and Ricci derivatives to know the isometry group and further analysis is required. 
\end{itemize} 
Here we study the more degenerate Ricci tensor, that is, Segre type [(1,111)].


\section{Isometries in type N vacuum metrics with cosmological constant}
\label{sec-N-lambda}

From now on, we consider vacuum solutions with cosmological constant $\Lambda$, that is, $Q = \lambda g$, $\Lambda = 6 \lambda$. In this case the Ricci tensor does not define any scalar or tensorial invariant, and consequently, the study of the symmetries depend on the sole Weyl tensor.


\subsection{Classes ${\rm C}n$, $n \leqslant 7$, compatible with $\, 6\,Q= \Lambda g$}
\label{lambda_Cn}

When $6\,Q= \Lambda g$, the Cotton tensor ${\cal C}$ vanishes, and Bianchi identities (\ref{BI_0}) imply that $\nabla_{\ell} {\cal L}$ is a null bivector, that is, $\varphi_1 =0$. Moreover, if $\ell \wedge a = \ell \wedge a^* =0$, then $(\ell, b) =(\ell, c)=0$, that is, $\varphi_1 = \varphi_2= 0$ implies that $\varphi_3 = \varphi_4= 0$. On the other hand, from (\ref{f-f*-Q}) we obtain $f = \mu \ell$, $f^* = \mu^* \ell$, which imply $\varphi_6=0$. Consequently, we obtain:
\begin{proposition} \label{Lambda_Cn}
In type N solutions with $6\,Q= \Lambda g$, the classes {\em C1}, {\em C3}, {\em C4} and {\em C6} given in definition {\em \ref{def-Cn}} are empty. 
\end{proposition}
The type N solutions with $6\,Q= \Lambda g$ that belong to classes {\rm C2}, {\rm C5} and {\rm C7} have a trivial isotropy group and the dimension of the isometry group can be obtained as stated in theorem \ref{theorem-Cn}. Therefore, what remains to be analysed are the isometries of the family $\widehat{\rm C7}$.


\subsection{Invariant scalars and constraints in the family ${\rm \widehat{C5}}$ with $6\,Q= \Lambda g$}
\label{C^5}

In the family ${\rm \widehat{C5}}$, which includes the family $\widehat{\rm C7}$, we have the expressions:
\be
a = \alpha \ell, \quad a^* = \alpha^* \ell;   \qquad f = \mu \ell, \quad f^* = \mu^* \ell  ,
\ee
where $\alpha, \alpha^*$ are given in (\ref{alpha}) and $\mu, \mu^*$ can be obtained as:
\be \label{mu}
\mu \equiv \frac{(f,x)}{(\ell,x)} , \qquad  \mu^* \equiv \frac{(f^*,x)}{(\ell,x)} ,
\ee
$x$ being an arbitrary time-like vector.

On the other hand, in addition to the fundamental vector $\ell$, we have the invariant vectors $b$ and $c$ given in (\ref{b-w}) and (\ref{c-v}), which can be written as:
\be \label{b_c_beta_gamma}
b = \beta^0 \ell + \beta e_2 + \beta^* \e_3 , \quad  c = \gamma^0 \ell + \gamma e_2 + \gamma^* e_3 ,
\ee
where $\beta, \beta^*, \gamma, \gamma^*$ are invariant scalars that can be obtained as:
\be \label{beta_gamma}
\beta \equiv -\frac{L(b,x)}{(\ell,x)} , \quad  \beta^* \equiv \frac{*L(b,x)}{(\ell,x)} , \quad  \gamma \equiv -\frac{L(c,x)}{(\ell,x)} , \quad  \gamma^* \equiv \frac{*L(c,x)}{(\ell,x)} ,
\ee
$x$ being an arbitrary time-like vector.

The Bianchi identities (\ref{BI_0}) imply the following constraints:
\be \label{BI_a}
5 \alpha - 2(\gamma^* + \beta) =0, \quad 5 \alpha^* + 2(\gamma - \beta^*) =0 ,
\ee
which can be joined together in the following vectorial expression:
\be \label{BI_b}
5 m = 2 (b + c^{\perp}) + \pi \ell , \quad    m \equiv  \alpha e_2 + \alpha^* e_3,  \quad c^{\perp} \equiv  - \gamma e_3 +\gamma^* e_2 . 
\ee

Moreover, from (\ref{lambda_alpha}) we obtain:
\be \label{Lambda}
\Lambda = -3( \alpha^2 + \alpha^{*2})\leqslant 0 ,  \qquad  \alpha \dif \alpha + \alpha^* \dif \alpha^* =0 .
\ee
And now, (\ref{d-alpha}) and (\ref{db-dc}) become:
\be \label{dif_alpha}
\dif \alpha = \mu \ell + \alpha^* c, \quad \dif \alpha^* = \mu ^* \ell - \alpha c,
\ee
\be  \label{db-dc_2}
\dif b = \ell \wedge e, \quad \dif c =  \ell \wedge e^{\perp}, \qquad e = \mu e_2 + \mu^* e_3 , \quad e^{\perp} = \mu e_3 - \mu^* e_2 .
\ee
Finally, from (\ref{Lambda}) and (\ref{dif_alpha}) we obtain:
\be \label{alpha_mu}
\alpha \mu + \alpha^* \mu^* =0 , \qquad \alpha e^{\perp} = \mu^* m, \quad \alpha^* e^{\perp} = - \mu m. 
\ee


\subsection{Vacuum solutions ($\Lambda=0$) in the family ${\rm \widehat{C5}}$: pp-waves}
\label{lambda=0}

In the family ${\rm \widehat{C5}}$ the vacuum condition is equivalent to $\alpha = \alpha^*=0$ as a consequence of (\ref{Lambda}). In this case, the first equation in (\ref{1ee}) and (\ref{2ee-1}) become:
\be \label{pp_nabla-l}
\nabla \ell  = w \otimes \ell ,  \qquad \dif w = 0 ,
\ee
and a function $\psi$ exists such that $w = \dif \psi$. Then, $\hat{\ell} = e^{-\psi} \ell$ is a covariantly constant null vector, $\nabla \hat{\ell} =0$, and the solution is a vacuum pp-wave \cite{kramer, Ehlers-Kundt}.

Conversely, if the spacetime is a vacuum pp-wave, the Ricci identities imply that the covariantly constant vector determines the fundamental direction of a type N Weyl tensor, and the fundamental vector fulfils (\ref{pp_nabla-l}). Moreover, the solution belongs to the family ${\rm \widehat{C5}}$. The analysis of the conditions that characterise this family when $\alpha = \alpha^*=0$ leads to: 
\begin{proposition} \label{prop-pp-w}
The vacuum pp-waves are the vacuum type N solutions in the family ${\rm \widehat{C5}}$, which are characterised by the condition
\be \label{pp_waves}
 \nabla \ell \bar{\wedge} \ell = 0 .
\ee
The vacuum pp-waves with a trivial isotropy group belong to class {\em C7}, that is, they  fulfil $\ell \wedge b \wedge c \not=0$. Otherwise, we have the vacuum plane-waves, that can be characterised by the complementary condition $\ell \bar{\wedge} \nabla  \ell =0$.
\end{proposition}
The above proposition offers an IDEAL characterisation of the pp-waves that comes naturally from our approach to analysing the symmetries. It is worth remarking that an IDEAL labelling of these  geometries based on the Bel-Robinson tensor has been recently presented \cite{Igor-pp-w}.

The study of the isometry group of the pp-waves has been widely discussed \cite{kramer, Ehlers-Kundt, Sippel-Goenner-1986}. In subsection \ref{subsec-pp-waves} we compare these known results for the pp-waves with trivial isotropy group with our approach. Now we analyse the isometry group of the plane waves.

For the plane waves we have $\alpha =\alpha^*=0$, $\dif \ell = \dif b = \dif c=0$, and $b = b_\ell \ell$, $c = c_\ell \ell$, where now $b_\ell$ and $c_\ell$ are Weyl invariants that can be obtained as:
\be \label{b_ell-c_ell}
b_\ell \equiv \frac{(b,x)}{(\ell,x)} , \qquad c_\ell \equiv \frac{(c,x)}{(\ell,x)} ,
\ee
$x$ being an arbitrary time-like vector. The gradients $\dif b_\ell$ and $\dif c_\ell$ are collinear with $\ell$. Thus, we have an invariant vector $\ell$ and at most one independent invariant scalar. Then, we can state:

\begin{theorem} \label{theorem-S}
The type N vacuum solutions with $\Lambda=0$ in the family $\widehat{\rm C7}$ are the plane waves. They admit a non-trivial isotropy group of dimension two, and we can consider two classes:
\begin{itemize}
\item[-]
Class {\rm OS1}: $\dif b_\ell \not=0$ or $\dif c_\ell \not=0$. The metric admits a {\rm G}$_5$ on $3$-dimensional null orbits.
\item[-]
Class {\rm OS2}: $\dif b_\ell =\dif c_\ell =0$. The metric admits a {\rm G}$_6$ on $4$-dimensional orbits.
In this last class we can consider two subclasses: {\em OS2a} when $b_\ell \not=0$, and {\em OS2b} when $b_\ell =0$.
\end{itemize}
\end{theorem}
It is worth remarking that the class OS1 and the two subclasses OS2a and OS2b match with the three classes of plane waves considered in the literature \cite{kramer, Ehlers-Kundt}. This can be shown starting from their known canonical form and obtaining the Weyl invariants involved in the definition of our classes. Note that $b_\ell = 0$ is equivalent to $\nabla \ell =0$. Thus, the metrics in subclass OS2b are the only vacuum pp-waves having the fundamental vector $\ell$ covariantly constant. In the other pp-waves this property is fulfilled by a vector proportional to $\ell$. A Cartan-Karlhede approach to the plane waves can be found in \cite{Coley-2012}. 


\subsection{Analyses of the family $\widehat{\rm C7}$ with $\Lambda \not=0$}
\label{family-C^7}

After the above results, what remains for us to study is the family $\widehat{\rm C7}$ with $\Lambda <0$ or, equivalently, those solutions with $\alpha^2 + \alpha^{*2} \not=0$. In addition to the constraints presented in subsection \ref{C^5}, now $\varphi_7=0$, that is:
\be \label{ell_b_c}
\ell \wedge b \wedge c = 0  \qquad (\leftrightarrow \quad  \beta \gamma^* - \beta^* \gamma =0) .
\ee
Moreover, the integrability conditions of expressions (\ref{dif_alpha}) lead to:
\begin{equation}
\dif \mu = \mu^0 \ell + \mu^* c - \mu b  + 3 \mu m ,  \qquad 
\dif \mu^{*} = \mu^{*0} \ell - \mu c - \mu^* b  + 3 \mu^* m .
\end{equation}
From here, and (\ref{BI_b}) and (\ref{ell_b_c}) we obtain:
\begin{equation}
\ell \wedge c \wedge \dif \mu  = -\frac56 \mu \ell \wedge c \wedge c^ {\perp},  \qquad 
\ell \wedge c \wedge \dif \mu^*  = -\frac56 \mu^* \ell \wedge c \wedge c^ {\perp} .
\end{equation}
Then, when $\ell \wedge c \not=0$ and $\mu \not=0$ (respectively, $\mu^* \not=0$), the vectors $\{c, \dif \mu\}$ (respectively, $\{c, \dif \mu^*\}$)  fulfil the conditions of proposition \ref{prop-{b,c}_base} and a $R$-frame can be obtained.
Note that $\dif \mu = \dif \mu^* = 0$ implies $\mu = \mu^* =0$ as a consequence of (\ref{Lambda}), (\ref{dif_alpha}) and (\ref{alpha_mu}). Moreover, from (\ref{db-dc_2}) this condition is equivalent to $\dif b =0$. 

When $\ell \wedge c \not=0$ and $\dif b =0$, we can change $\{e_2, e_3\}$ in the adapted frame so that the vector $c$ given in (\ref{b_c_beta_gamma}) takes the expression $c = \gamma e_2 + \gamma^* e_3$. Moreover (\ref{db-dc_2}) and (\ref{d-alpha}) become $\dif b = \dif c = 0$, $\dif \alpha = \alpha^* c, \dif \alpha^* = - \alpha c$, and (\ref{1ee}), (\ref{BI_a}) and (\ref{ell_b_c}) imply:
\be \label{dif_ell_c}
\dif \ell \wedge c \not=0, \qquad  b = \psi_\ell \ell + \psi_c c, 
\ee
with $\beta = \psi_c \gamma$, $\beta^* = \psi_c \gamma^*$. The scalars $\psi_c$ and $\psi_\ell$ are Weyl invariants that can be obtained as:
\be \label{psi_ell_c}
\psi_c \equiv \frac{(b,c)}{c^2}, \qquad    \psi_{\ell} \equiv \frac{1}{(\ell,x)}(b-\psi_c c,x),  
\ee
where $x$ is an arbitrary time-like vector. From this expression we get $\dif \psi_c \wedge c + \dif \psi_\ell \wedge \ell + \psi_\ell \dif \ell =0$. Then, $(\dif \psi_c, \ell) = (\dif \psi_\ell , \ell)=0$, and $\dif \psi_\ell \wedge \ell \wedge c + \psi_\ell\, \dif \ell \wedge c =0$. From (\ref{dif_ell_c}), if $\psi_\ell \not=0$ the vectors $\{c, \dif \psi_\ell\}$ fulfil the conditions of proposition \ref{prop-{b,c}_base} and a $R$-frame can be obtained. 

Let us consider $\ell \wedge c \not=0$, $\dif b =0$ and $\psi_\ell =0$. Then,  $b = \psi_c c$ and $\dif \psi_c \wedge c=0$, and all the invariant scalars already considered are collinear with $c$. For instance, $\dif \gamma = (\alpha + \frac15 \gamma^*) c$, $\dif \gamma^* = (\alpha^* - \frac15 \gamma) c$. Then, we can calculate the covariant derivative of $c = \gamma e_2 + \gamma^* e_3$ and, taking into account that $\dif c = 0$, we obtain:
\begin{equation}
\nabla c = \chi \ell \otimes \ell - (\gamma \alpha + \gamma^* \alpha^*) g  +\frac45  \psi_c c \otimes c  ,
\end{equation}
$\chi$ being a scalar invariant that can obtained as:
\be \label{chi} 
\chi \equiv \frac{P(x, x)}{(\ell, x)^2} , \qquad P \equiv \nabla c + (\gamma \alpha + \gamma^* \alpha^*) g  -\frac45 \psi_c c \otimes c   ,
\end{equation}
where $x$ is an arbitrary time-like vector. The study of the Ricci identities for the vector $c$ implies $(\dif \chi, \ell)=0$ and the equivalence of the three following conditions:
\be \label{dif_chi}
\dif \chi \wedge \ell \wedge c = 0 \quad \leftrightarrow \quad \chi = \frac{5 \gamma \gamma^*}{3\, c^2} \quad  \leftrightarrow \quad \dif \chi \wedge c = 0 .
\ee
Thus, if these conditions do not hold the vectors $\{c, \dif \chi\}$ fulfil the conditions of proposition \ref{prop-{b,c}_base} and a $R$-frame can be obtained. When conditions (\ref{dif_chi}) hold, there is only one independent invariant scalar and two independent invariant vectors $\ell$ and $c$, and an isometry group G$_4$ on time-like 3-dimensional orbits exists.

Let us suppose now that $\ell \wedge c =0$ ($\gamma = \gamma^*=0$). Then, necessarily, $\ell \wedge b \not =0$ (otherwise, (\ref{BI_a}) implies $\alpha = \alpha^* =0$). Now, we can change $\{e_2, e_3\}$ in the adapted frame so that the vector $b$ given in (\ref{b_c_beta_gamma}) becomes $b =  \frac52 (\alpha e_2 + \alpha^* e_3)$, where we have made use of the Bianchi identity (\ref{BI_a}). If we calculate its covariant derivative  taking into account (\ref{dif_alpha}) and (\ref{1ee}), we obtain:
\begin{equation}
\nabla b = - \frac52 (\alpha^2+ \alpha^{*2}) g + \frac25 b \otimes b + \frac12 \phi \, \ell \otimes \ell + \frac32 e \otimes \ell + \frac52 \ell \otimes e ,
\end{equation}
where $e = \mu e_2 + \mu^* e_3$, and $\phi$ is a scalar function. Then, taking its symmetric part we arrive to:
\begin{equation} \label{S}
\hspace{-15mm} S = \ell \, \tilde{\otimes} \, q , \qquad q= \phi \ell + 4 \, e, \quad  
S \equiv \nabla b + ^t\! \nabla b +5 (\alpha^2+ \alpha^{*2}) g - \frac45 b \otimes b  .
\end{equation}
Then, if $\mu^2+\mu^{*2} \not=0$ ($\dif b \not=0$), we have that $q$ is an invariant vector which can be obtained as:
\be \label{q}
q \equiv \frac{1}{(\ell,x)^2} [(\ell, x) S(x) - \frac12 S(x,x) \ell] .
\ee
Moreover, $q$ is orthogonal to $b$ as a consequence of (\ref{BI_b}) and (\ref{alpha_mu}), and the vectors $\{b, q\}$  fulfil the conditions of proposition \ref{prop-{b,c}_base} and a $R$-frame can be obtained.

When $\ell \wedge c =0$ and $\dif b =0$, we have that $q = \phi \ell$, and $\phi$ is a Weyl invariant that can be obtained as:
\be \label{phi}
\phi \equiv \frac{(q,x)}{(\ell,x)} ,
\ee
where $x$ is an arbitrary time-like vector. Moreover, $\, 2 \nabla b = \frac45 b \otimes b - 5 (\alpha^2+ \alpha^{*2}) g + \phi \ell \otimes \ell$. Then, the Ricci identities for the vector $b$ lead to: 
\be \label{d_phi}
\ell \wedge (\dif \phi + \frac25  b + 5 \tilde{m} ) = 0, \qquad  \tilde{m} = \alpha e_2 - \alpha^* e_3 .
\ee
Then, $\ell \wedge b \wedge \dif \phi \not=0$ if, and only if, $\alpha \alpha^* \not=0$. When this happens, the vectors $\{b, \dif \phi\}$ fulfil the conditions of proposition \ref{prop-{b,c}_base} and a $R$-frame can be obtained. 

Let us consider $\ell \wedge c =0$, $\dif b =0$, and $\alpha \alpha^*=0$. If $\alpha^*=0$ (respectively, $\alpha =0$), then, $c =0$ and $\dif \alpha =0$ (respectively $\dif \alpha^* =0$) as a consequence of (\ref{dif_alpha}), and $\tilde{m}= \alpha e_2 = \frac25 b$ (respectively $\tilde{m}= -\alpha^* e_3 = -\frac25 b$). Moreover, among all the invariant scalars already considered, only $\phi$ is, generically, non-constant. If $\dif \phi =0$, all the scalar invariants are constant and only two independent invariant vectors, $\ell$ and $b$, exist. Then,  an isometry group G$_5$ on O$_4$ exists. If $\dif \phi \not=0$, (\ref{d_phi}) implies $\dif \phi = \phi_{\ell} \ell + 2 (\varepsilon + \frac15\phi) b$, where $\phi_{\ell}$ is an invariant scalar that can be obtained as:
\be \label{phi_ell_b}
\phi_{\ell} \equiv \frac{1}{(\ell,x)}(\dif \phi-\phi_b b,x), \qquad \phi_b \equiv -2(\varepsilon + \frac15\phi) , 
\ee
where $x$ is an arbitrary time-like vector, and $\varepsilon=1$ if $\alpha^*=0$ and $\varepsilon=-1$ if $\alpha=0$. Note that $\dif \phi_{\ell} \wedge \ell \wedge b = 0$. Thus, there are at most two independent scalars. Thus, we have two cases. If $\dif \phi \wedge \dif \phi_\ell = 0$ a G$_4$ on time-like 3-dimensional orbits exists. Finally, when $\dif \phi \wedge \dif \phi_\ell \not= 0$, a G$_3$ on null 2-dimensional orbits exists.


\subsection{Summary: classification and characterisation theorem for the family $\widehat{\rm C7}$}
\label{Summary_Lambda}

The analysis presented above induces a classification of the type N vacuum solutions (with $\Lambda$) in the family $\widehat{\rm C7}$. Now we introduce the definition and notation of each class.
\begin{definition} \label{def-Kn} 
Let ${\cal L}$ and $\ell$ be the fundamental bivector and the fundamental vector of a type {\em N} Weyl tensor. Let us consider the Weyl concomitants introduced in definition {\em \ref{def-Cn}} and $\alpha$ and $\alpha^*$ given in {\em (\ref{alpha})}, $b$ given in {\em (\ref{B})} and {\em (\ref{b-w})}, $c$ given in {\em (\ref{C})} and {\em (\ref{c-v})}, $\gamma$ and $\gamma^*$ given in {\em (\ref{beta_gamma})}, $\psi_\ell$ given in {\em (\ref{psi_ell_c})}, $\chi$ given in {\em (\ref{chi})}, $\phi$ and $q$ given in {\em (\ref{phi})}, {\em (\ref{q})} and {\em (\ref{S})}, and $\phi_\ell$ given in {\em (\ref{phi_ell_b})}. 

In the family $\widehat{\rm C7}$ ($\varphi_n = 0, n \leqslant 7$) of type N vacuum solutions with non-vanishing cosmological constant we consider the following classes defined by invariant conditions:
\[\begin{array}{lllll}
\hspace{-20mm}{\rm AR1} &\quad  \ell \wedge c \not=0  &\quad    \dif b \not=0   &\quad     & \\
\hspace{-20mm}{\rm AR2} &\quad \ell \wedge c \not=0  &\quad   \dif b  =0  &\quad   \psi_\ell \not=0 & \\
\hspace{-20mm}{\rm AR3} &\quad  \ell \wedge c \not=0  &\quad    \dif b  =0  &\quad   \psi_\ell =0  &\quad   3 \chi c^2  \not= 5 \gamma \gamma^* \\
\hspace{-20mm}{\rm AS} &\quad  \ell \wedge c \not=0 &\quad    \dif b  =0  &\quad   \psi_\ell =0  &\quad   3 \chi c^2  = 5 \gamma \gamma^* \\
\hspace{-20mm}{\rm BR1} &\quad  \ell \wedge c =0  &\quad   \dif b \not=0  &\quad    &  \\
\hspace{-20mm}{\rm BR2} &\quad  \ell \wedge c =0  &\quad    \dif b  =0  &\quad   \alpha \alpha^* \not=0 &  \\
\hspace{-20mm}{\rm BS1} &\quad  \ell \wedge c =0  &\quad    \dif b  =0  &\quad   \alpha \alpha^* =0  &\quad   \dif \phi \wedge \dif \phi_\ell \not=0 \\
\hspace{-20mm}{\rm BS2} &\quad  \ell \wedge c =0  &\quad    \dif b  =0  &\quad   \alpha \alpha^* =0  &\quad   \dif \phi \wedge \dif \phi_\ell =0,  \qquad  \dif \phi \not=0, \\
\hspace{-20mm}{\rm BS3} &\quad  \ell \wedge c =0  &\quad    \dif b  =0  &\quad   \alpha \alpha^* =0  &\quad   \dif \phi =0 
\end{array}\]
\end{definition}
The invariants $b$, $c$, $\gamma$, $\gamma^*$, $\alpha$, $\alpha^*$ and $\psi_\ell$ are of first-order in the Weyl tensor, while $\chi$, and $\phi$ are of second-order, and $\Phi_\ell$ is of third-order.  
\begin{theorem} \label{theorem-A_B}
All the type N vacuum solutions with non-vanishing cosmological constant in the classes {\rm AR}n, $n=1,2,3$, and {\rm BR}p, $p=1,2$, are regular, that is, they admit a $R$-frame ${\cal B} =  \{\ell, k, e_2, e_3\}$, which can be obtained as stated in proposition {\em \ref{prop-{b,c}_base}} with:
\begin{itemize}
\item[-]
In classs {\rm AR1}, $\{b, c\} = \{c, \dif \mu\}$ if $\mu \not=0$ or $\{b, c\} = \{c, \dif \mu^*\}$ if $\mu^* \not=0$.
\item[-]
In classs {\rm AR2}, $\{b, c\} = \{c, \dif \psi_\ell\}$.
\item[-]
In classs {\rm AR3}, $\{b, c\} = \{c, \dif \chi\}$.
\item[-]
In classs {\rm BR1}, $\{b, c\} = \{b,q\}$.
\item[-]
In classs {\rm BR2}, $\{b, c\} = \{b,\dif \phi\}$.
\end{itemize}
If $H$ is the connection tensor {\em (\ref{defH_null})} associated  with ${\cal B}$, the dimension of the isometry group is obtained as stated in the algorithm given in figure {\em 1} of {\em \cite{SMF-ST}}.
\end{theorem}
The scalar invariants $\mu, \mu^*$ used in class AR1 are given in (\ref{mu}) and (\ref{f_f*}), and the invariant vector $q$ used in class BR1 is given in (\ref{q}) and (\ref{S}).

\begin{theorem} \label{theorem-AB-singular}
All the type N vacuum solutions with non-vanishing cosmological constant in the classes {\rm AS}, and {\rm BS}p, $p=1,2,3$, are singular, that is, they admit a non-trivial isotropy group:
\begin{itemize}
\item[-]
In class {\rm AS} the metric admits a {\rm G}$_4$ on $3$-dimensional time-like orbits.
\item[-]
In class {\rm BS1} the metric admits a {\rm G}$_3$ on $2$-dimensional null orbits.
\item[-]
In class {\rm BS2} the metric admits a {\rm G}$_4$ on $3$-dimensional time-like orbits.
\item[-]
In class {\rm BS3} the metric admits a {\rm G}$_5$ on $4$-dimensional orbits.
\end{itemize}
\end{theorem}

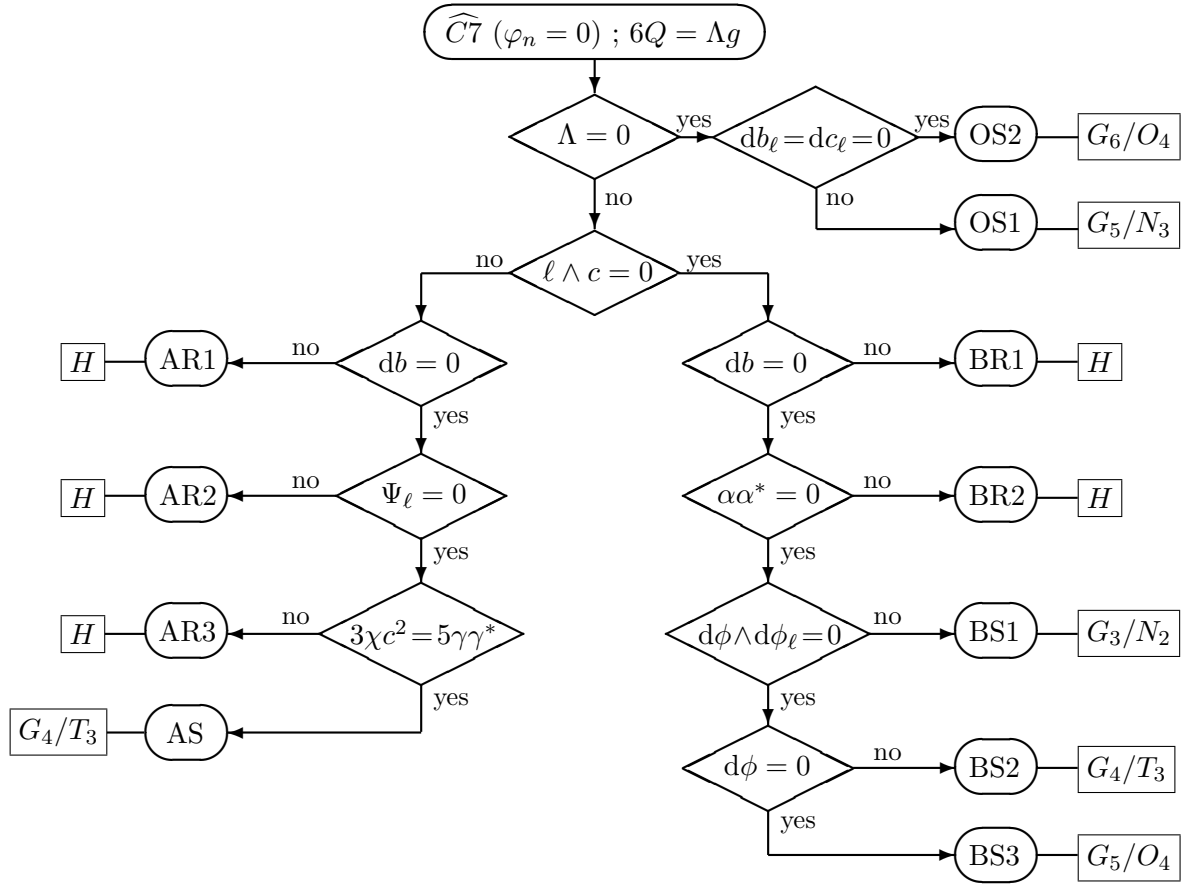
\begin{figure}[b]
\vspace*{0.3cm}
 \setlength{\unitlength}{0.9cm} {\small \noindent

\begin{picture}(1,13)


\thicklines

\put(10,11.5){\footnotesize{yes}}
\put(13.5,11.5){\footnotesize{yes}}

\put(8.95,10.35){\footnotesize{no}}
\put(12.2,10.35){\footnotesize{no}}

\put(10.15,9.45){\footnotesize{yes}}

\put(7.05,9.45){\footnotesize{no}}

\put(12.75,8.15){\footnotesize{no}}
\put(4.35,8.15){\footnotesize{no}}

\put(12.75,6.2){\footnotesize{no}} \put(4.35,6.2){\footnotesize{no}}

\put(12.9,4.2){\footnotesize{no}} \put(4.2,4.2){\footnotesize{no}}
\put(12.9,2.2){\footnotesize{no}}

\put(11.53,7.15){\footnotesize{yes}}
\put(6.43,7.15){\footnotesize{yes}}

\put(11.53,5.15){\footnotesize{yes}}
\put(6.43,5.15){\footnotesize{yes}}

\put(11.53,3){\footnotesize{yes}} \put(6.43,3){\footnotesize{yes}}

\put(11.53,1.2){\footnotesize{yes}}


\put(8.8,12.5){\vector(0,-1){0.5}}

\put(8.8,10.75){\vector(0,-1){0.75}}

\put(10,9.35){\line(1,0){1.35}} \put(7.6,9.35){\line(-1,0){1.35}}

\put(11.35,9.35){\vector(0,-1){0.7}}

\put(6.25,9.35){\vector(0,-1){0.7}}

\put(5,8.03){\vector(-1,0){1.6}} \put(2.2,8.05){\line(-1,0){0.6}}

\put(6.25,7.4){\vector(0,-1){0.7}}

\put(11.35,7.4){\vector(0,-1){0.7}}

\put(6.25,5.47){\vector(0,-1){0.67}}

\put(11.35,5.47){\vector(0,-1){0.67}}

\put(11.35,3.3){\vector(0,-1){0.59}}

\put(11.35,1.45){\line(0,-1){0.65}}

\put(11.35,0.8){\vector(1,0){2.75}}

\put(15.3,0.8){\line(1,0){0.6}}

\put(12.6,2.08){\vector(1,0){1.51}} \put(15.3,2.08){\line(1,0){0.6}}

\put(12.8,4.05){\vector(1,0){1.31}} \put(15.3,4.05){\line(1,0){0.6}}

\put(4.8,4.05){\vector(-1,0){1.4}}

\put(2.2,4.05){\line(-1,0){0.6}}

\put(6.25,3.3){\line(0,-1){0.69}}

\put(6.25,2.6){\vector(-1,0){2.85}}
\put(2.18,2.6){\line(-1,0){0.55}}

\put(12.6,6.08){\vector(1,0){1.5}}

\put(5,6.08){\vector(-1,0){1.6}} \put(2.2,6.08){\line(-1,0){0.6}}


\put(8.8,12.9){{\oval(5,0.8)}} \put(6.6,12.75){$\widehat{C7}$
($\varphi_n =0 $) ;  $6 Q = \Lambda g$ }


\put(8.8,11.98){\line(-2,-1){1.25}}
\put(8.8,11.98){\line(2,-1){1.25}} \put(8.8,10.73){\line(2,1){1.25}}
\put(8.8,10.73){\line(-2,1){1.25}} \put(8.25,11.25 ){$\Lambda=0$}

\put(12.05,12.1){\line(-2,-1){1.5}}
\put(12.05,12.1){\line(2,-1){1.5}} \put(12.05,10.6){\line(2,1){1.5}}
\put(12.05,10.6){\line(-2,1){1.5}} \put(10.94,11.2){$\dif b_\ell\! =
\!\dif c_\ell\! =\!0$} \put(13.5,11.35){\vector(1,0){0.6}}

 \put(15.28,11.35){\line(1,0){0.63}}

\put(14.7,11.4){{\oval(1.2,0.8)}} \put(14.3,11.25){OS2}

\put(15.9,11.25){\framebox{$G_6/O_4$}}

 \put(10,11.35){\vector(1,0){0.54}}

\put(14.7,10.1){{\oval(1.2,0.8)}} \put(14.3,9.95){OS1}
\put(15.9,9.95){\framebox{$G_5/N_3$}}

 \put(15.28,10){\line(1,0){0.63}}
 \put(12.06,10){\vector(1,0){2.05}}
\put(12.06,10.6){\line(0,-1){0.6}}


\put(8.8,9.98){\line(-2,-1){1.25}} \put(8.8,9.98){\line(2,-1){1.25}}
\put(8.8,8.73){\line(2,1){1.2}} \put(8.8,8.73){\line(-2,1){1.25}}
\put(8.04,9.23 ){$\ell \wedge c=0$}


\put(11.35,8.65){\line(-2,-1){1.25}}
\put(11.35,8.65){\line(2,-1){1.25}}
\put(11.35,7.4){\line(2,1){1.25}} \put(11.35,7.4){\line(-2,1){1.25}}
\put(10.72,7.9){$\dif b =0$}

\put(12.6,8.03){\vector(1,0){1.5}}

\put(14.7,8.1){{\oval(1.2,0.8)}} \put(14.3,7.95){BR1}
\put(15.9,7.9){\framebox{$H$}}

\put(15.3,8.05){\line(1,0){0.6}}

\put(6.25,8.65){\line(-2,-1){1.25}}
\put(6.25,8.65){\line(2,-1){1.25}} \put(6.25,7.4){\line(2,1){1.25}}
\put(6.25,7.4){\line(-2,1){1.25}} \put(5.62,7.9){$\dif b =0$}

\put(2.8,8.1){{\oval(1.2,0.8)}} \put(2.4,7.95){AR1}
\put(0.95,7.9){\framebox{$H$}}



\put(6.25,6.7){\line(-2,-1){1.25}} \put(6.25,6.7){\line(2,-1){1.25}}
\put(6.25,5.45){\line(2,1){1.25}} \put(6.25,5.45){\line(-2,1){1.25}}
\put(5.65,5.97){$\Psi_\ell=0$}

\put(11.35,6.7){\line(-2,-1){1.25}}
\put(11.35,6.7){\line(2,-1){1.25}}
\put(11.35,5.45){\line(2,1){1.25}}
\put(11.35,5.45){\line(-2,1){1.25}} \put(10.6,5.97){$\alpha
\alpha^*=0$}

\put(2.8,6.1){{\oval(1.2,0.8)}} \put(2.4,5.95){AR2}
\put(0.95,5.9){\framebox{$H$}}

\put(14.7,6.1){{\oval(1.2,0.8)}} \put(14.3,5.95){BR2}
\put(15.9,5.9){\framebox{$H$}} \put(15.3,6.05){\line(1,0){0.6}}


\put(11.35,4.8){\line(-2,-1){1.5}} \put(11.35,4.8){\line(2,-1){1.5}}
\put(11.35,3.3){\line(2,1){1.5}} \put(11.35,3.3){\line(-2,1){1.5}}
\put(10.32,3.9){$\dif \phi\! \wedge\! \dif \phi_\ell\!=\!0$}

\put(14.7,4.1){{\oval(1.2,0.8)}} \put(14.3,3.95){BS1}

\put(15.9,3.95){\framebox{$G_3/N_2$}}

\put(14.7,2.1){{\oval(1.2,0.8)}} \put(14.3,1.95){BS2}

\put(15.9,1.95){\framebox{$G_4/T_3$}}

\put(14.7,0.8){{\oval(1.2,0.8)}} \put(14.3,0.65){BS3}

\put(15.9,0.65){\framebox{$G_5/O_4$}}

\put(6.25,4.8){\line(-2,-1){1.5}} \put(6.25,4.8){\line(2,-1){1.5}}
\put(6.25,3.3){\line(2,1){1.5}} \put(6.25,3.3){\line(-2,1){1.5}}
\put(5.2,3.9){$3\chi c^2\!=\!5 \gamma \gamma^*$}

\put(2.8,4.1){{\oval(1.2,0.8)}} \put(2.39,3.95){AR3}
\put(0.95,3.9){\framebox{$H$}}


\put(2.8,2.6){{\oval(1.2,0.8)}} \put(2.5,2.45){AS}
\put(0.2,2.5){\framebox{$G_4/T_3$}}


\put(11.35,2.7){\line(-2,-1){1.25}}
\put(11.35,2.7){\line(2,-1){1.25}}
\put(11.35,1.45){\line(2,1){1.25}}
\put(11.35,1.45){\line(-2,1){1.25}} \put(10.7,1.97){$\dif \phi=0$}

\end{picture} }
\vspace*{-0.5cm} 
\caption{This flow diagram distinguishes the different classes in the family $\widehat{\rm C7}$, and gives de dimension of the group and orbits for the singular ones.} \label{figure-2}
\end{figure}


%

The IDEAL nature of these results enables us to perform an algorithm that we present as a flow diagram in figure \ref{figure-2}. It offers the dimension of the isometry group for the vacuum solutions with cosmological constant $\Lambda$ in the family $\widehat{\rm C7}$. The input data are the Weyl concomitants given in definition \ref{def-Kn}. The case $\Lambda=0$ leads to the plane waves, which have two singular classes. When $\Lambda \not=0$, we have the two branches A and B, both containing regular and singular classes. In the singular ones the end arrows indicate the dimension of the group and orbits, and in the regular ones the end arrows lead to the determination of the associated connection tensor, to which we must apply the algorithm given in \cite{SMF-ST}.

\section{Discussion, examples and work in progress}
\label{sec-discussion}

In this paper we have obtained an IDEAL approach to the isometries of a type N vacuum solution with cosmological constant. This study offers the following insight: given a metric of this family in an arbitrary coordinate system, $g_{\alpha \beta}(x^\gamma)$, we can determine, following an algorithmic procedure, the dimension of its ismometry group and the dimension of its orbits. 

In our approach. we also analyse the symmetries of a family of type N metrics (independently of the Ricci tensor) which includes the twisting metrics, the Robinson-Trautman solutions and a wide family of the regular Kundt metrics. 

Our study induces an invariant classification of the type N metrics, which is based on constraints on the Weyl concomitants that enable us to build a $R$-frame, when it exists. We analyse the relation of our classes with the more usual classifications based on the optic coefficients associated with the Weyl fundamental direction.

Several direct consequences easily follow from our study. For example, we can ask ourselves when the fundamental vector determines a Killing direction. This is an open question, but we have the following partial result: \\[1mm]
Let $\ell$ the fundamental vector of a Petrov-Bel type N metric. Then:
\begin{itemize}
\item[-]
In the family $\widehat{\rm C5}$, $\ell$ is collinear with a Killing vector if, and only if, $\dif b =0$. When this holds, no space-like orbits exist.
\item[-]
Except for classes {\rm AR1} and {\rm BR1}, in all the vacuum solutions (with $\Lambda$) in the family $\widehat{\rm C7}$, $\ell$ is collinear with a Killing vector. 
\item[-]
In the classes C1 and C2 (and in particular in the twisting and Robertson-Trautman metrics) $\ell$ is never collinear with a Killing vector.
\end{itemize}

On the other hand, from the results summarised in theorem \ref{theorem-Cn}, we have:\\[1mm]
The following Petrov-Bel type N metrics admit a $R$-frame and, consequently, have a trivial isotropy group:
\begin{itemize}
\item[-]
The solutions with non-geodesic fundamental direction (belong to class {\rm C1}).
\item[-]
The solutions with twisting geodesic fundamental direction (belong to class {\rm C2}).
\item[-]
The Robertson-Trautman metrics (belong to class {\rm C2}).
\item[-]
The vacuum solutions with positive cosmological constant (belong to class {\rm Cn}, $n\leqslant 7$).
\end{itemize}
Furthermore, we have the following result about vacuum solutions :
\begin{itemize}
\item[-]
The only type N vacuum solutions ($\Lambda =0$) with non-trivial isotropy group are the plane waves.
\item[-]
The type {\em N} vacuum solutions ($\Lambda =0$) with trivial isotropy group either are the regular pp-waves or they belong to classes {\rm C}n, $n\leqslant 5$.
\end{itemize}
%
%



\subsection{Implementing the algorithms on \textit{xAct}}
\label{subsec-xAct}
The algorithmic structure of the IDEAL approach obtained in the previous sections makes it especially well-suited for implementation in a formal computational framework. An example of such a formal calculation program is \textit{xAct}, a collection of Mathematica packages designed for tensor computer algebra \cite{xAct}. One of the authors of this paper (SM) is developing, in collaboration with A. García-Parrado, an \textit{xAct} package called \textit{xIdeal} that implements IDEAL characterisations and other algorithms devised by our group. The implementation of the algorithms given in figures \ref{figure-1} and \ref{figure-2}, and many others, in \textit{xIdeal} functions that automatise their application is a work in progress that will be published elsewhere when finished. However, a preliminary version can be found at \cite{xIdeal}. 

Still, in the following subsections, we will use \textit{xIdeal} to help us apply the algorithms presented in this paper to meaningful families of solutions as examples: the pp-waves, the Kundt waves and the Siklos metrics. For the three of them, we will obtain which class or classes Cn they belong to by applying the algorithm in figure \ref{figure-1}. Once the class is determined, for the pp-waves and the Kundt waves we will determine a $R$-frame as stated in theorem \ref{theorem-Cn} and obtain the corresponding connection tensors. After that, we could apply the \textit{xIdeal} function that implements the algorithm given in figure 1 of \cite{SMF-ST} to determine the dimension of the isometry group for the most general cases. However, such \textit{xIdeal} function works faster for expressions where the dependence of the metric functions on the space-time coordinates is more explicit. Thus, for the pp-waves we will apply it to known particular subfamilies, while for the Kundt waves we will simply prove, with the help of \textit{xIdeal}, that they admit, at most, a G$_2$ as it is already known. For the Siklos solutions, instead, we will apply the algorithm in figure \ref{figure-2} to the particular solutions belonging to the $\widehat{\rm C7}$ family. All these computations can be found in the \textit{Mathematica} notebook attached as supplementary material.


\subsection{The pp-waves with trivial isotropy group}
\label{subsec-pp-waves}

For vacuum and Maxwell null or radiation fields, the metric of the pp-waves takes the expression \cite{kramer}:
\be \label{metric-pp}
\dif s^2 = \dif x^2 + \dif y^2 - 2 \dif u \dif v - 2h(u,x,y) \dif u^2.
\ee
In the vacuum case, $h(u,x,y) = f(u, \zeta) + \bar{f}(u, \bar{\zeta})$, where $f$ is an analytical complex function and $\zeta = x + \ci y$, that is, $h_{xx} + h_{yy} =0$. The symmetries of the vacuum pp-waves can be found in the Ehlers and Kundt review \cite{Ehlers-Kundt} (see also \cite{kramer}), and the study for the non-vacuum case was achieved in \cite{Sippel-Goenner-1986}. 

Invariant analyses of the vacuum pp-waves have been addressed with different methods. It is worth mentioning the broad Cartan-Karlhede approach by Milson  {\it et al.} \cite{Coley-JMP}, the scalar differential invariants procedure \cite{ Kruglikov}, or the recent IDEAL characterisation by Khavkine {\it et al.} \cite{Igor-pp-w}. Here, we apply our algorithms and compare our results with the already known ones. 

If we apply the algorithm given in figure \ref{figure-1} to the metric (\ref{metric-pp}) we obtain that, generically, it belongs to class C7 (see the attached supplementary \textit{Mathematica} notebook for the explicit computations). Only in a degenerate subclass we have $\varphi_7 = 0$, which leads to the plane waves whose symmetries have already been considered in theorem \ref{theorem-S}. The expression of the metric function $f(u, \zeta)$ for these degenerate cases is known \cite{kramer}, and our algorithm in figure \ref{figure-2} offers their intrinsic labelling: (i) if $f(u, \zeta)= A(u) \zeta^2$, then we obtain class OS1; (ii) if $f(u, \zeta)= u^{2(\ci\kappa -1)} \zeta^2$, then we obtain class OS2a; (iii) if $f(u, \zeta)= a e^{2 \ci\kappa u} \zeta^2$, then we obtain class OS2b. 

Now, we focus on the pp-waves in class C7, that is, those with a trivial isotropy group. We can determine a $R$-frame as stated in theorem \ref{theorem-Cn} and obtain the corresponding connection tensor $H_7$. The \textit{xIdeal} function that performs the algorithm given in figure 1 of \cite{SMF-ST} enables us to determine the symmetries of a specific pp-wave solution. Note that the pp-waves fulfil $\dif b = 0$, and thus there is a Killing vector collinear with $\ell$. Consequently, at least a G$_1$ exists and there are no space-like orbits.

The expression of the function $f(u, \zeta)$ corresponding to the different isometry groups is known \cite{kramer, Ehlers-Kundt}. We apply our approach to each of these metrics and the results are summarised in table \ref{table-1}. It is worth mentioning that, for the metrics in rows 6 and 7, we actually applied our approach to two particular cases. In the column 'invariant conditions' we find, on the left, the condition (equations G$_3$ or G$_{2a}$) in the algorithm given in figure 1 of \cite{SMF-ST} that each metric fulfils, which determines the dimension of the group. On the right, we find the invariant conditions that complete the properties of the group. 
\begin{table}[t]
\vspace*{-2mm}
\caption{This table summarises the analysis of the regular pp-waves by usisng our IDEAL approach.}
\label{table-1}
\vspace*{-3mm}
$$\begin{array}{llcll}
 \hline \hline\\[-5mm] 
f(u, \zeta) &\quad   {\rm Invariant}&\!\!\!{\rm conditions}    &  \quad   {\rm Group} \ {\rm G}_r   & \  {\rm O}_r \  {\rm Orbits} 
\\ \hline \hline 

e^{\ci \alpha} \zeta^{2 \ci k}  &\quad   {\rm Eq.} \  {\rm G_3} &   Z = 0 &\quad   {\rm G_3}\quad {\rm BI} & \ {\rm T}_3 \ \ [\tilde{m}^2 \!>\!0]  \\ 

e^{2 k \zeta} &\quad   {\rm Eq.} \   {\rm G_3} &  {\cal B}_1(Z)  &\quad   {\rm G_3}\quad {\rm BIII} &\ {\rm T}_3 \ \ [\tilde{m}^2 \!>\!0]    \\ 

\ln \zeta  &\quad   {\rm Eq.} \    {\rm G_3} &   {\cal B}_2(Z)   &\quad   {\rm G_3}\quad {\rm BVI_{0}} &\ {\rm T}_3 \ \ [\tilde{m}^2 \!>\!0]    \\

a u^{-2} \ln \zeta  &\quad   {\rm Eq.} \   {\rm G_3} &   {\cal C}_2(Z)   &\quad   {\rm G_3}\quad {\rm BVI_{0}} & \ {\rm T}_3 \ \ [\tilde{m}^2 \!>\!0]    \\

A(u) \ln \zeta &\quad    {\rm Eq.} \    {\rm G_{2a}} &  (A, H) \! = \! 0    &\quad   {\rm G_2}\quad \textsf{Abelian} &  \ {\rm N}_2 \ \ [(\!A, \! A\!)\!=\!0]     \\ 

f(\zeta e^{\ci k u}) &\quad   {\rm Eq.} \  {\rm G_{2a}} &   (A, H) \! = \! 0    &\quad   {\rm G_2}\quad \textsf{Abelian}&    \  {\rm T}_2 \  \  [(\!A, \! A\!)\!<\!0]          \\ 

u^{-2}A(\zeta u^{\ci k}) &\quad   {\rm Eq.} \  {\rm G_{2a}} &(A, H) \! \not= \! 0    &\quad   {\rm G_2}\quad \textsf{non-Abelian} & \  {\rm  T}_2  \ \  [(\!A, \! A\!)\!<\!0]        
\\ 

f(\zeta,u) &\quad   
&\quad     &\quad   {\rm G_1}\quad  &  \ {\rm N}_1       \\
 \hline \hline 
\end{array}$$
\vspace{-4mm}
\end{table}

For the G$_3$, we can determine the Bianchi type as pointed out in \cite{SMF-ST}. For any vector $v$ and any 2-form $V$, the vector $\tilde{m}_\lambda = C^{[1]}_{\lambda \rho \mu \nu}v^\rho V^{\mu \nu} $, defines the (space-like) direction orthogonal to the orbits. Then, the induced metric on the orbits is $\tilde{\gamma} = g - n \otimes n$, with $n = |\tilde{m}^2|^{-1/2} \tilde{m}$. Moreover, 
\begin{equation} \label{Z}
Z_{\alpha \beta} = - \frac{1}{2}\tilde{\gamma}^{\lambda}_{\alpha} \,
{H_{\lambda}}^{\mu \nu} \eta_{\mu \nu \beta \rho} \, n^{\rho} .
\end{equation}
is the structure tensor associated to the isometry group, and we can apply the algorithm to determine the Bianchi type presented in \cite{FS-L3}. In table \ref{table-1}, ${\cal B}_1(Z)$ and ${\cal B}_2(Z)$ denote the invariant conditions (given in \cite{FS-L3}) that lead to Bianchi types BIII and BVI$_0$, respectively. The two classes with this last Bianchi type can be discriminated by the invariant condition $b^2 = 0$ ($b^2 \not=0$), where $b$ is given in (\ref{b-w}). The invariant $\tilde{m}$ is a space-like vector and, therefore, the orbits are time-like.

For the G$_2$, the connection tensor fulfils equations G$_{2a}$. Then the commutative character can be expressed with the invariant condition $(A, H) = 0$, where $(A,H)$ denotes the contraction of the 2-form $A$ with the two first indices of $H$, and $A$ is $A_{\alpha \beta} = \eta(C^{[1]}, C^{[1]})_{\alpha \beta \lambda \mu \nu \rho \sigma \tau} v_1^\lambda V_1^{\mu \nu} v_2^\rho V_2^{\sigma \tau}$, $v_1, v_2$ being two arbitrary vectors and $V_1, V_2$ two arbitrary 2-forms \cite{SMF-ST}. Moreover, the causal character of the orbits depends on the sign of $(A,A)$.

For any other function $f(\zeta, u)$, we know that, at least, a G$_1$ exists. Thus, necessarily one of the conditions G$_{1a}$, G$_{1b}$, G$_{1c}$ or G$_{1d}$ of the algorithm  given in figure 1 of \cite{SMF-ST} holds. This fact can be tested by considering specific choices of the function $f$ not belonging to the previously considered subfamilies.


\subsection{On the Kundt waves}
\label{subsec-Kundt-w}

In his original paper, Kundt \cite{Kundt} considers a family of vacuum and Maxwell null or radiation field solutions whose metric line element takes the expression \cite{Coley-2013, Kruglikov}:
\be \label{metric-Kundtwaves}
\dif s^2 = \dif x^2 + \dif y^2 -  \dif u \dif v +  \frac{2v}{x} \dif u \dif x - [8 x h(u,x,y)-\frac{v^2}{4x^2}] \dif u^2.
\ee
In the vacuum case $h(u,x,y)$ fulfils the partial differential equation $h_{xx} + h_{yy} = 0$. These solutions have been analysed using the Cartan-Karlhede and the scalar differential invariant approaches  \cite{Coley-2013, Kruglikov}. Moreover, it is known that they can admit two, one or no symmetries \cite{Coley-2013}.

Here, we do not follow our approach to studying the symmetries in detail. However, we want to comment on two direct results (the explicit derivations can be found in the supplementary \textit{Mathematica} notebook). On the one hand, if we apply our algorithm given in figure \ref{figure-1} to the metric (\ref{metric-Kundtwaves}) we obtain $\varphi_5 = - (4 x^4)^{-1} \not=0$, and it belongs to class C5. Consequently, all the Kundt waves have a trivial isotropy group, and the study of their isometry group should be addressed using the connection tensor $H$ associated with the $R$-frame obtained as stated in theorem \ref{theorem-Cn}.

On the other hand, a simple reasoning shows that the Kundt waves admit, at most, a G$_2$. Indeed, $\varphi_5$ is a non-constant scalar invariant. If either $\alpha$ or $\alpha^*$ (given in (\ref{alpha})) determine a new independent invariant scalar, then we have, at most, a $G_2$. Otherwise, we have that $\dif \alpha \wedge \dif \varphi_5 = \dif \alpha^* \wedge \dif \varphi_5 = 0$. But these conditions lead to $h = \alpha_1(u) e^{k_1 y} \sin(k_1x + k_2) + \alpha_2(u) x + \alpha_3(u) y + \alpha_4(u)$, with $k_i$ arbitrary constants, and if we apply the \textit{xIdeal} function that performs the algorithm  given in figure 1 of \cite{SMF-ST} to such metric, we obtain that, generically, it has no symmetries. Moreover, for this metric  $\eta(C^{[1]},C^{[1]}) \not=0$, and consequently a G$_3$ cannot exist. If we take $\alpha_2(u) = \alpha_3(u) = \alpha_4(u) =0$ and $\alpha_1(u) = k_0$, the \textit{xIdeal} function determines that a G$_2$ exists, in agreement with the results in \cite{Kruglikov}.


\subsection{On the Siklos solutions}
\label{subsec-Siklos}
 
The Siklos metrics \cite{Siklos} are vacuum and Maxwell null or radiation field solutions with negative cosmological constant $\Lambda$. Their metric line element takes the expression \cite{Siklos, Podolsky-98}:
\be \label{metric-Siklos}
\dif s^2 = -\frac{\Lambda}{x^2}[\dif x^2 + \dif y^2 - 2 \dif u \dif v - 2 h(u,x,y) \dif u^2].
\ee
In the vacuum case $h(u,x,y)$ fulfils the partial differential equation $h_{xx} + h_{yy} = 2 h_x/x$. These solutions were analysed in \cite{Podolsky-98}, and their symmetries were studied in the Siklos paper \cite{Siklos} and recently revisited in \cite{Calvaruso}. These metrics turn out to be a subclass of the family of Kundt metrics obtained in \cite{Garcia-Plebanski} (see also \cite{Ortaggio}). 

If we apply the algorithm given in figure \ref{figure-1} to the metric (\ref{metric-Siklos}) we obtain that, in general, it belongs to class C7. However, some particular subfamilies belong to the family $\widehat{\rm C7}$ (see the attached supplementary \textit{Mathematica} notebook for the explicit computations). 

For the general case, belonging to class C7, we can determine a $R$-frame as stated in theorem \ref{theorem-Cn} and obtain the corresponding connection tensor $H_7$. The \textit{xIdeal} function that performs the algorithm given in figure 1 of \cite{SMF-ST} enables us to determine the symmetries of a specific Siklos solution. Note that in this case $\dif b = 0$, and thus there is a Killing vector collinear with $\ell$. Consequently, at least a G$_1$ exists and there are no space-like orbits. A detailed analysis, such as that done for pp-waves in the previous subsection, requires further work that goes beyond the scope of this paper. Now, we only point out that the cases 0, 1, 2, 3, 5 and 6 considered in \cite{Calvaruso} have $\varphi_7 \not=0$ and belong to class C7. 

The application of the algorithm given in figure \ref{figure-2} to the cases in the family $\widehat{\rm C7}$ shows that only our singular classes in the branches A and B are possible: cases 4, 7 ($\equiv$ 11), 9 and 10 in \cite{Calvaruso} correspond to our classes BS1, BS3, BS2 and AS, respectively.


\subsection{Ending comments}
\label{ending}

The IDEAL approach presented in this paper determines the dimensions of the isometry group and of its orbits for a type N vacuum solution with cosmological constant. However, this study also opens new questions that require further work. 


On the one hand, our approach should be extended to other energy contents, and especially to perfect fluid and null or non-null Einstein-Maxwell solutions. We have shown here that the regular classes C1, C3, C4 and C5 are not compatible with the vacuum condition. Then, a further question is to study the Ricci tensors that are compatible with these classes.

On the other hand, we propose addressing this type of study for the case of solutions of Petrov-Bel types D or O, where the Weyl tensor does not algebraically determine a $R$-frame either.


\ack 
We thank A Garc\'ia-Parrado for sharing his knowledge about \textit{xAct}. This work has been supported by the Generalitat Valenciana Project CIAICO/2022/252 and the Plan Recuperaci\'on, Transformaci\'on y Resiliencia, project ASFAE/2022/001, with funding from European Union NextGenerationEU (PRTR-C17.I1). S.M. acknowledges financial support from the Generalitat Valenciana (grant CIACIF/2021/028).

\section*{Data availability statement}
All data that support the findings of this study are included within the article (and any supplementary files).

\appendix

\section{Notation}
\label{apendix-A}

The following conventions have been followed in the text:
\begin{description}
\item $(x,y)\equiv g(x,y)= g_{\alpha \beta} x^{\alpha} y^{\beta}$,
$x^2 =(x,x)$ for vectors $x$, $y$ and a metric tensor $g$.
\item $B(x)_{\alpha} = B_{\alpha \beta } x^{\beta}$, $B(x,y)= B_{\alpha \beta }
x^{\alpha} y^{\beta}$, for a 2-tensor $B$ and vectors $x$, $y$.
\item $(A^2)_{\alpha \mu} =A_{\alpha \beta} {A^{\beta}}_{\mu}$, for
an arbitrary 2-tensor $A$.
\item $(T \cdot x)_{\bar{p}} = T_{\bar{p} \alpha} x^{\alpha}$, for an
arbitrary tensor $T$ and a vector $x$, $\bar{p}$ denoting
multi-index.

\item $(v \wedge w)_{\alpha \beta} = v_\alpha w_\beta - w_{\alpha}
v_{\beta}$ for arbitrary 1-forms $v$, $w$.

\item $(A \bar{\wedge} v)_{\bar{p} \beta \gamma} = A_{\bar{p} \beta}
v_{\gamma}- A_{\bar{p} \gamma} v_{\beta}$, $(v \bar{\wedge}
A)_{\alpha \beta \bar{p}} = v_{\alpha} A_{\beta \bar{p}} -
v_{\beta} A_{\alpha \bar{p}}$, $A$ being an arbitrary tensor and $v$
a 1-form, $\bar{p}$ denoting multi-index.

\item $({\cal W}^{2})_{\alpha \beta \mu \nu} = \frac{1}{2} {\cal
W}_{\alpha \beta \rho \sigma} \, {{\cal W}^{\rho \sigma}}_{\mu \nu}$
for a double bivector ${\cal W}$.

\item ${\cal W}({\cal X})_{\alpha \beta} = \frac{1}{2}
 {\cal W}_{\alpha \beta \mu \nu} {\cal X}^{\mu \nu}$, for a double
 bivector ${\cal W}$ and a bivector ${\cal X}$.

 \item ${\cal W}({\cal X}, {\cal X}) = \frac{1}{4}
 {\cal W}_{\alpha \beta \mu \nu} {\cal X}^{\alpha \beta} {\cal X}^{\mu \nu}$, for a double
 bivector ${\cal W}$ and a bivector ${\cal X}$.

\item $(*F)_{\alpha \beta} = \frac12 \eta_{\alpha \beta \lambda \mu} F^{\lambda \mu}$, for a 2-form $F$. 

\end{description}

\section{Structure equations in type N spacetimes}
\label{apendix-B}

Let $\{\ell, k, e_2, e_3\}$ be an adapted frame of a type N spacetime, $L= \ell \wedge e_2$ the fundamental 2-form, and $U= \ell \wedge k$. If Ric is the Ricci tensor, let us consider $2 Q = \Ric - \frac{\r}{6} g $.
\\[1mm]
\noindent
{\it First structure equations}
\begin{equation}  \label{1ee}
\begin{array}{l}
\nabla \ell = w \otimes \ell + a \otimes e_2 + a^{*} \otimes e_3 \\
\nabla k = - w \otimes k + s \otimes e_2 + s^* \otimes
e_3 \\
\nabla e_2 = a \otimes k + s \otimes \ell + v \otimes e_3 \\
\nabla e_3 = a^* \otimes k + s^* \otimes \ell - v \otimes e_2
\end{array}
\end{equation}

\noindent
{\it Second structure equations}
\begin{eqnarray} 
\dif w  - a\wedge s - a^* \wedge s^* = Q(\ell) \wedge k + \ell
\wedge Q(k) \label{2ee-1}  \\
\dif a  + a \wedge w + a^* \wedge v = Q(e_2) \wedge \ell + e_2
\wedge Q(\ell) \label{2ee-2}  \\
\dif a^* + a^* \wedge w - a \wedge v = Q(e_3) \wedge \ell + e_3
\wedge Q(\ell) \label{2ee-3}  \\
\dif v + a^* \wedge s + s^* \wedge a = Q(e_3) \wedge e_2 + e_3
\wedge Q(e_2) \label{2ee-4}  \\
\dif s - s \wedge w - v \wedge s^* = Q(e_2) \wedge k + e_2 \wedge
Q(k) +  \ell \wedge e_2  \label{2ee-5}  \\
\dif s^* -s^* \wedge w + v \wedge s = Q(e_3) \wedge k + e_3 \wedge
Q(k) + e_3 \wedge \ell  \label{2ee-6}
\end{eqnarray}
\noindent
{\it Covariant derivative of the fundamental bivector}
\be \label{nablaL}
\nabla {\cal L} = (w - \ci v) \otimes {\cal L} + (a - \ci a^*) \otimes {\cal U}
\ee

\noindent
{\em Bianchi identities} state that $\nabla \cdot W = {\cal C}$, where ${\cal C}$ is the Cotton tensor, ${\cal C}_{\alpha \beta \gamma}  = \nabla_{\alpha} Q_{\beta \gamma} - \nabla_{\beta} Q_{\alpha \gamma}$. For a type N spacetime, and as a consequence of (\ref{Weyl-N}) and (\ref{nablaL}), they take the expression (when ${\cal C} =0$):
\be \label{BI_0}
2 {\cal L} (w - \ci v) + {\cal U}(a - \ci a^*) = 0 , \qquad {\cal L}(a - \ci a^*) = 0 .
\ee
%


\end{document}